\documentclass[sigconf,9pt]{acmart}
\usepackage[ruled,noend]{algorithm2e}
\usepackage[english]{babel}
\usepackage{blindtext}
\usepackage{hyperref}
\usepackage{xcolor}
\usepackage{url}

\usepackage{subfigure}
\usepackage{diagbox}
\usepackage{caption}
\usepackage{multirow}

\usepackage{pifont}
\usepackage{algorithm2e}

\acmYear{2024}\copyrightyear{2024}
\setcopyright{rightsretained}
\acmConference[ACM SIGCOMM '24]{ACM SIGCOMM 2024 Conference}{August 4--8, 2024}{Sydney, NSW, Australia}
\acmBooktitle{ACM SIGCOMM 2024 Conference (ACM SIGCOMM '24), August 4--8, 2024, Sydney, NSW, Australia}
\acmDOI{10.1145/3651890.3672222}
\acmISBN{979-8-4007-0614-1/24/08}

\begin{document}
\title{NegotiaToR: Towards A Simple Yet Effective On-demand Reconfigurable Datacenter Network}

\author{Cong Liang\textsuperscript{$1$}, Xiangli Song\textsuperscript{$1$}, Jing Cheng\textsuperscript{$1$}, Mowei Wang\textsuperscript{$2$}, Yashe Liu\textsuperscript{$2$}, \\Zhenhua Liu\textsuperscript{$2$}, Shizhen Zhao\textsuperscript{$3$}, Yong Cui\textsuperscript{$1$}}

\affiliation{
  \institution{\textsuperscript{$1$}Tsinghua University \hspace{8pt} \textsuperscript{$2$}Huawei Technologies Co., Ltd \hspace{8pt} \textsuperscript{$3$}Shanghai Jiao Tong University}
  \city{}
  \state{}
  \country{}
}

\renewcommand{\shortauthors}{Liang et al.}

\begin{abstract}

Recent advances in fast optical switching technology show promise in meeting the high goodput and low latency requirements of datacenter networks (DCN). We present NegotiaToR, a simple network architecture for optical reconfigurable DCNs that utilizes on-demand scheduling to handle dynamic traffic. In NegotiaToR, racks exchange scheduling messages through an in-band control plane and distributedly calculate non-conflicting paths from binary traffic demand information. Optimized for incasts, it also provides opportunities to bypass scheduling delays. NegotiaToR is compatible with prevalent flat topologies, and is tailored towards a minimalist design for on-demand reconfigurable DCNs, enhancing practicality. Through large-scale simulations, we show that NegotiaToR achieves both small mice flow completion time (FCT) and high goodput on two representative flat topologies, especially under heavy loads. Particularly, the FCT of mice flows is one to two orders of magnitude better than the state-of-the-art traffic-oblivious reconfigurable DCN design.

\end{abstract}



\renewcommand{\authors}{Cong Liang, Xiangli Song, Jing Cheng, Mowei Wang, Yashe Liu, Zhenhua Liu, Shizhen Zhao, Yong Cui}
\maketitle

\section{Introduction}

With more and more applications running in the cloud, traffic demands in DCNs increase continuously \cite{jupiter-rising}. Among the applications, tasks like high performance computing (HPC) are greatly affected by both goodput and latency, putting stringent requirements on the network \cite{10.5555/3388242.3388246}. 
However, existing packet-switched DCNs struggle to meet these requirements due to inadequate capacity of switching chips \cite{sirius} which will worsen with the slowdown of Moore's Law \cite{moore, 8727133}. 
To address this, the development of optical switching technology \cite{6301681, 928320, 6691912, 9125762, 8535334, 9489945, 10117390, 8535333} especially fast optical switching \cite{9125762, 8535334, 9489945, 8535333} has led researchers to turn to reconfigurable optical DCNs \cite{helios, c-through, mordia, OSA, firefly, reactor, projector, rotornet, opera, sirius, trod, 10.1145/3491054, 9772727, jupiter-evolving, topoopt, tpuv4, PULSE}, which provide higher capacity as well as lower cost compared with packet switching. 

Unlike packet-switched networks that rely on buffers to absorb conflicts, bufferless optical switching requires synchronous scheduling and reconfiguration to accommodate the dynamic traffic demands including incasts \cite{meta2022imc, dctcp, 10.1145/3131365.3131375, 10.1145/2079296.2079304}.
Recently, traffic-oblivious reconfigurable DCNs \cite{rotornet, shoal, sirius} have drawn researchers' attention because of their simplicity. They schedule the network according to predefined rules and use data relay to cope with dynamic traffic patterns. 
However, despite their simplicity, the data-relay design can lead to compromised performance in goodput and latency. 
Consequently, there is a growing need for a new design that can effectively handle dynamic traffic demands, ensuring both high goodput and low latency while still maintaining feasibility.

We present NegotiaToR, an on-demand reconfigurable DCN architecture with a simple design. With in-band distributed scheduling, it dynamically adapts the optical links among top-of-rack (ToR) switches to traffic demands, and sends data directly to destinations through one-hop paths. Beyond the scheduled connections, NegotiaToR also provides unscheduled connections, mitigating the impact of scheduling delays even under incasts. Utilizing existing arrayed waveguide grating routers (AWGR) and fast-tunable lasers, NegotiaToR is compatible with prevalent flat topologies, achieving a better performance than the state-of-the-art traffic-oblivious scheme on the same hardware with similar complexity.

On-demand is an intuitive solution to handle the dynamic traffic in DCNs, especially for unpredictable ToR-ToR traffic where demand forecasting based on historical data is difficult \cite{10.1145/2079296.2079304}. When put into the context of fast optical switching, previous on-demand solutions \cite{PULSE, 
helios, c-through, mordia, OSA, reactor, firefly} introduce excessive scheduling complexity and raise practicality concerns.
The key challenge of NegotiaToR thus lies in realizing scalable on-demand scheduling with high practicality.

First of all, scheduling often leads to reduced practicality due to the scale of DCNs \cite{DBLP:conf/sigcomm/RoyZBPS15} and the dynamics of traffic demands \cite{meta2022imc, dctcp, 10.1145/3131365.3131375, 10.1145/2079296.2079304}. Measuring detailed traffic demands of all senders and then using them as input to calculate non-conflicting paths often introduce high complexity. In contrast, NegotiaToR's demand information is binary. No data size information is needed by the scheduling algorithm. Meanwhile, NegotiaToR distributes the scheduling to ToRs, where each ToR only accounts for the scheduling of ingoing and outgoing traffic, achieving a comparable low complexity with traffic-oblivious designs.

Then, it is usually expensive to build and maintain a robust control plane for distributed scheduling in on-demand optical reconfigurable DCNs. 
Supporting a distributed scheduling algorithm requires ToRs to frequently transmit scheduling messages while avoiding conflicts with data transmission. An independent control network can solve the problem, but will introduce additional deployment and maintenance costs.
NegotiaToR implements an in-band control plane, where all ToRs periodically establish all-to-all connectivity with fast optical switching. The scheduling algorithm runs on it in a pipelined manner. Senders can access the scheduling results locally and tune lasers to the derived wavelength without the need to deliver scheduling results.

Finally, ensuring low latency of mice flows, particularly when incast happens, is challenging in scheduling. 
Mice flows that occupy a large number of flows in DCNs \cite{homa} are latency-sensitive and usually arrive simultaneously as incasts due to the partition/aggregate design pattern \cite{dctcp} of DCN applications. 
Other than lowering the scheduling delay, NegotiaToR utilizes the periodical all-to-all connectivity to piggyback a small volume of mice flow data along with scheduling messages, providing opportunities to bypass scheduling delays. This effectively manages incasts, promising application performance even under concurrent traffic demands.

The design of NegotiaToR is guided by the principle of Occam's Razor. For deployment practicality, we hope to find a minimalist design of on-demand reconfigurable DCNs. Possibilities are also explored to trade off simplicity for better performance. 
We evaluate NegotiaToR through simulations on two representative flat topologies. 
Results show that NegotiaToR outperforms the state-of-the-art traffic-oblivious reconfigurable DCN design under comparable complexity, both in FCT and goodput. Particularly, NegotiaToR's mice flow FCT is one to two orders of magnitude better.

To summarize, we make the following contributions:
\begin{itemize}
    \item We present NegotiaToR, a simple reconfigurable DCN architecture with scalable on-demand scheduling for flat topologies with fast optical switching enabled \textbf{(\S\ref{sec-design-all})}. It comprises the distributed NegtotiaToR Matching scheduling algorithm \textbf{(\S\ref{sec:negotiator-matching})}, the two-phase epoch serving as in-band control plane and data plane \textbf{(\S\ref{sec:negotiator_network_stack})}, and a mechanism to bypass scheduling delays even under incasts \textbf{(\S\ref{sec:bypass_scheduling_delay})}.
    \item We explore the possibilities to trade off simplicity for better performance, and show that extra complexity does not necessarily translate into
    proportionate performance gains \textbf{(\S\ref{sec:algorithm_design_discussion})}.
    \item We evaluate NegotiaToR through simulations (\textbf{\S\ref{sec:evaluation}}). Results show that with similar complexity, NegotiaToR outperforms the state-of-the-art traffic-oblivious design in both FCT and goodput.
\end{itemize}

\textit{This work does not raise any ethical issues.}

\section{Background \& Motivation}
\label{sec:motivation}

Optical switching technology has developed rapidly, and there has been a great interest in using them in DCNs in recent years \cite{10.1145/3491054, helios, c-through, mordia, OSA, reactor, firefly, projector, rotornet, opera, sirius, trod, 9772727, jupiter-evolving, topoopt, 10117390, tpuv4}. 
To fully utilize the high capacity and power efficiency of optical switching, one trend is to connect racks with optical switches directly \cite{OSA, projector, rotornet, opera, sirius, topoopt}. This poses two main considerations for optical switching hardware: supporting the large number of ToRs, and accommodating the highly dynamic traffic demands among them. 

\begin{figure}[t]
    \centering
        \centering
        \subfigure[{Parallel network topology built with high port-count AWGRs}]{
        \includegraphics[width=0.20\textwidth]{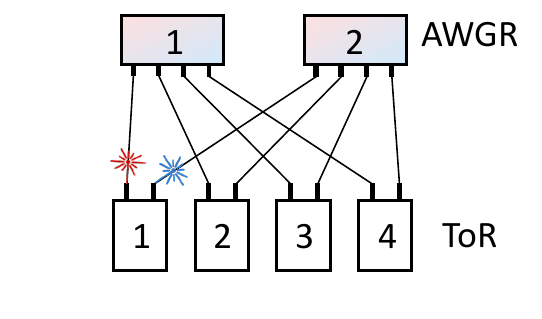}
        \label{fig:negotiator_topology_big_switch}
    }
    \hfill
        \subfigure[{Thin-clos topology built with low port-count AWGRs}]{
        \includegraphics[width=0.20\textwidth]{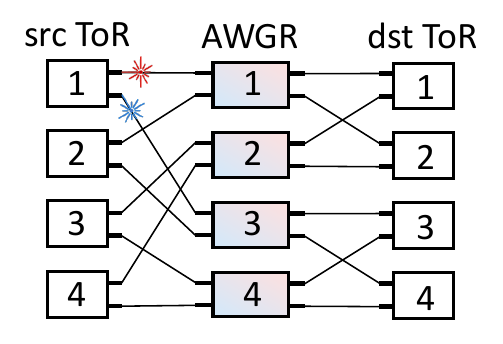}
        \label{fig:negotiator_topology_thin_clos}
    }

    \caption{Flat topologies in AWGR-based optical DCNs.}
    \label{fig:negotiator_topology}
\end{figure}

\vspace{6pt}

\noindent{\textbf{Fast optical switching technology is ready for DCNs.}} Recent advances in AWGR-based switching, mainly faster reconfiguration (e.g., nanoseconds end-to-end reconfiguration delay \cite{sirius}) and higher port count (e.g., 270 \cite{6301681}, 400 \cite{928320}, and 512 \cite{6691912} ports), have enabled it to meet the dynamic traffic demands of DCN. AWGR is a fully-passive optical switch. By tuning the wavelength of the tunable laser at the source, data can be forwarded to different destinations. This makes it a fit for distributed scheduling, where the source can tune the wavelength according to the scheduling results derived locally.
Flat topologies with one layer of AWGRs shown in Figure \ref{fig:negotiator_topology} are commonly used, due to the insertion loss and wavelength limitation of AWGR cascading \cite{6552916, 6634216}.

Both high and low port-count AWGRs are capable of connecting all ToRs in the DCN scale.
Notably, for high port-count AWGRs, a single AWGR can interconnect all ToRs, facilitating the implementation of the parallel network topology as shown in Figure \ref{fig:negotiator_topology_big_switch}.
For more accessible low port count AWGRs that are insufficient to connect all ToRs alone, the thin-clos topology \cite{6552916, 6634216} illustrated in Figure \ref{fig:negotiator_topology_thin_clos} becomes a practical alternative. 
In this topology, unlike the former one, each port of a ToR can only connect to a subset of other ToRs, and all ports together can reach the whole network.
Meanwhile, recent advancements in fast-tunable lasers \cite{8976124, sirius, 9333367, 9125762, 8535334, 9489945, 8535333} and time synchronization \cite{211255, sirius} have reduced the end-to-end reconfiguration delay to as low as 10 nanoseconds \cite{sirius}, which is sufficient to cope with the dynamic traffic demands among ToRs.

\vspace{6pt}

\noindent{\textbf{Scheduling optical interconnections among ToRs with dynamic and unpredictable traffic demands is challenging.}} 
For optical interconnections at higher switch levels, like the spine layer, since the traffic pattern is typically predictable due to job placement strategies, infrequent reconfiguration based on historical trace and traffic engineering techniques are sufficient \cite{trod, jupiter-evolving} for performance.

However, regarding inter-ToR connections, they face the challenge of highly dynamic and unpredictable traffic which includes bursty scenarios like incasts \cite{meta2022imc, dctcp, 10.1145/3131365.3131375, 10.1145/2079296.2079304}. 
The development of fast switching technology has made it possible to reconfigure the network on-demand to adapt to the dynamic traffic. 
A centralized scheduler can do the job, but faces practicality concerns because of the scheduler's limited scalability.
Recently, researchers have turned to traffic-oblivious designs \cite{shoal, sirius}. The network reconfigures itself regularly in a round-robin manner to provide all-to-all connectivity, regardless of actual traffic patterns. To mitigate the resulting mismatch of network connectivity and traffic demands, they use Valiant's Load Balancing (VLB) \cite{vlb, 10.1016/S0140-3664(01)00427-3} to adapt the traffic to the network, uniforming the traffic pattern to all-to-all by spreading data across the network before routing it to the final destination, and thus utilizing all links. Such approaches are practical but come at the expense of goodput and latency. 
Data relay doubles the traffic volume, competing for receivers' bandwidth, potentially causing congestion and damaging goodput where worst-case goodput can downgrade to 50\%. 
Meanwhile, the detouring also damages mice flow FCT, particularly when elephant flows are spread across the network and block the mice ones at intermediate nodes.
The performance downgrade worsens under heavier loads, which is a critical concern for HPC tasks like large-scale ML training where large amounts of flows are synchronously released to the network \cite{9772727, 10.5555/3388242.3388246, hoefler2023datacenter}.
NegotiaToR is designed to meet these needs, offering a practical solution that can accommodate the high-performance requirements of modern DCNs where fast optical switching technology is ready.

\section{NegotiaToR Design}
\label{sec-design-all}

NegotiaToR is a simple network architecture for reconfigurable DCNs where bufferless optical links connect buffered ToRs. With minimal traffic demand information, it schedules traffic distributedly on the ToRs via the in-band control plane in an on-demand while scalable manner.

\subsection{Design overview}
\label{sec-design}

NegotiaToR is compatible with prevalent flat topologies.
As depicted in Figure \ref{fig:negotiator_topology}, we choose two representative flat topologies to demonstrate our design, the parallel network topology that necessitates high port-count AWGRs, and the thin-clos topology that only needs readily available low port-count AWGRs. 
In both topologies, ToRs' uplink ports are equipped with fast-tunable lasers and attached to AWGR-based optical switches. One ToR maintains a FIFO queue for each of the other ToRs in the network. Data are first sent to this per-destination queue before being put into per-port queues and heading for their destinations. For data transmission, ToRs are time-synchronized\footnote{There is no need for AWGRs to be time synchronized since they are entirely passive.}, and simultaneously send bits according to non-conflicting port-level matches that the topology can provide.

\begin{figure}[t]
    \centering
        \includegraphics[width=0.43\textwidth]{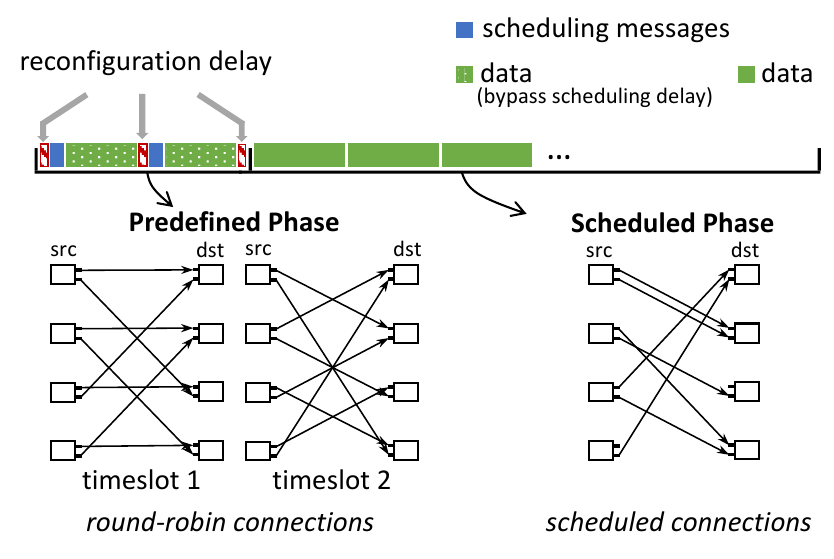}
    \caption{Each epoch in NegotiaToR comprises two phases, and reconfiguration only happens in the predefined phase.}
    \label{fig:negotiator_epoch}
\end{figure}

To this end, NegotiaToR employs a distributed scheduling mechanism on a per-epoch basis, where an epoch is a fixed-length time interval. Each epoch comprises two phases, working as the in-band control plane and data plane \textbf{(\S\ref{sec:negotiator_network_stack})}. 
All packets are transmitted directly to their destinations through one-hop paths. We present the design of one epoch in Figure \ref{fig:negotiator_epoch}. 
In the predefined phase, ToRs exchange scheduling messages through the all-to-all connectivity provided by round-robin via fast wavelength switching, enabling distributed scheduling. Note that network reconfiguration only happens in this phase, lowering reconfiguration overhead.
In the subsequent scheduled phase, ToRs set the wavelengths to the locally derived scheduling results and send data packets. To mitigate the impact of scheduling delays, especially under incasts, NegotiaToR also piggybacks a small volume of data with scheduling messages utilizing the unscheduled connections in the predefined phase, bypassing the scheduling delay especially for latency-sensitive mice flows \textbf{(\S\ref{sec:bypass_scheduling_delay})}.

\begin{figure*}[t]
    \centering
        \centering
        \begin{minipage}{0.75\textwidth}
        \subfigure[An example on the parallel network topology. REQUEST: sources send binary requests representing traffic demands to their destinations. GRANT: conflict elimination at the destinations' side (i.e., many-to-one). ACCEPT: conflict elimination at the sources' side (i.e., one-to-many).
        The accepted grants indicate a set of non-conflicting matches]{
        \includegraphics[width=\textwidth]{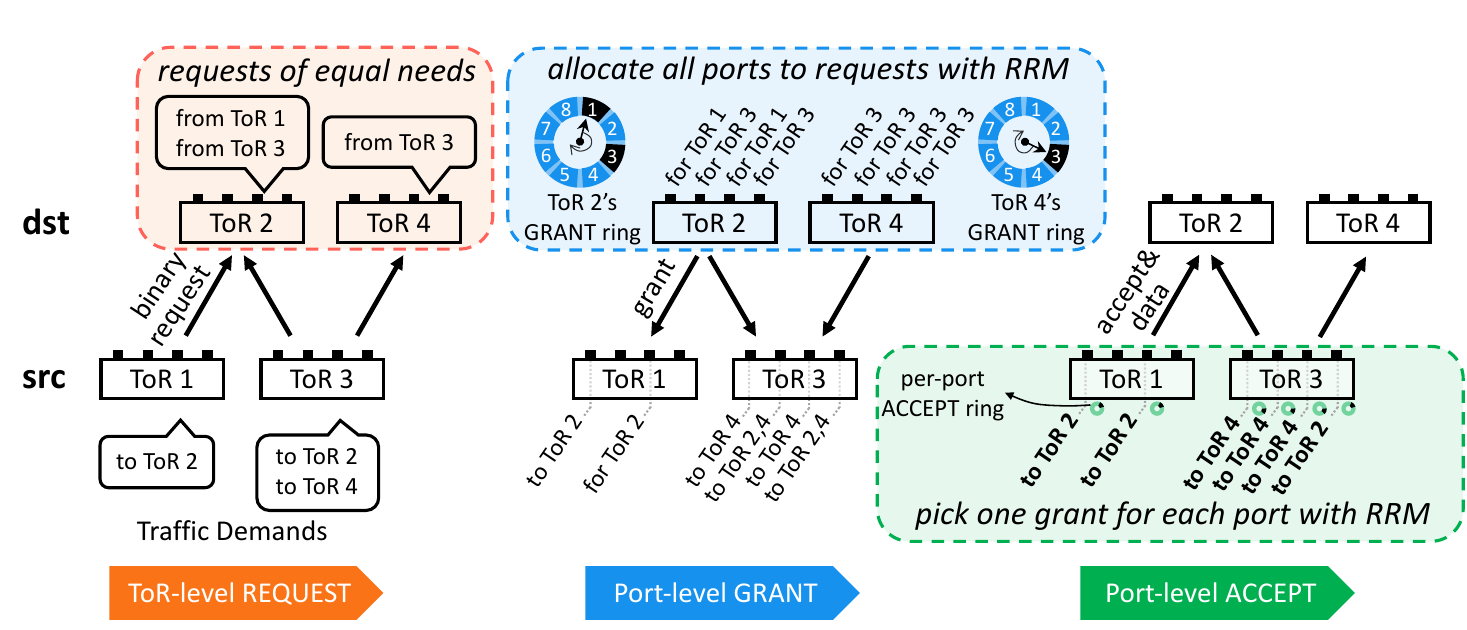}
        \label{fig:the_algorithm}
    }
    \end{minipage}
    \hfill
    \begin{minipage}{0.2\textwidth}
        \centering
        \subfigure[Per-ToR GRANT ring on the parallel network topology]{
        \includegraphics[width=\textwidth]{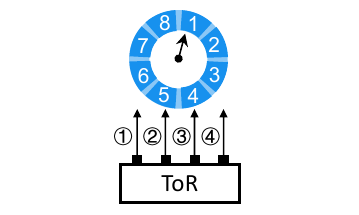}
        \label{fig:the_grant_parallel_network}
    }
    \vfill
    \subfigure[Per-port GRANT rings on the thin-clos topology]{
        \includegraphics[width=\textwidth]{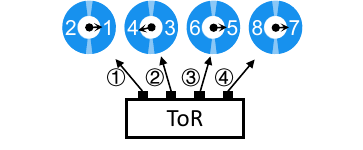}
        \label{fig:the_grant_thin_clos}
    }
    \end{minipage}
    \caption{NegotiaToR Matching's workflow. (a) shows the workflow on the parallel network topology. (b) and (c) illustrate the variance in the GRANT step across two topologies, attributable to differing connection capabilities. 
    On the parallel network topology, port \textcircled{1}-\textcircled{4}'s grant priority is determined by a shared ring, whereas on thin-clos it's determined by port-specific rings.
    Once granted, the pointer is updated to prioritize the next source.}
    \label{fig:negotiator_matching_algorithm}
\end{figure*}

Utilizing the in-band control plane, ToRs distributedly run a simple on-demand scheduling algorithm, NegotiaToR Matching \textbf{(\S\ref{sec:negotiator-matching})}, which accounts for both network performance and scalability. 
The algorithm does not require traffic forecasting and utilizes a non-iterative approach, making it suitable for the DCN scenario where ToR-ToR traffic is highly dynamic and the round-trip time (RTT) between ToRs is often long (e.g., several microseconds) compared with nanoseconds reconfiguration delay.
Consequently, iterative distributed scheduling would result in long scheduling delays, adversely affecting mice flow performance that could otherwise benefit from fast optical switching.
Each ToR functions like a negotiator. ToRs get simplified binary local traffic demands from their per-destination queues, exchange scheduling messages regularly with each other, and make scheduling decisions locally according to their incoming and outgoing traffic needs. A new scheduling process starts at the beginning of each epoch. One process lasts for three epochs, and the scheduling for consecutive epochs is performed in a pipelined manner. 
This way, NegotiaToR distributes on-demand scheduling computation to each ToR and adapts the network to real-time traffic, realizing feasibility and high performance.

Guided by the principle of Occam's Razor, we carefully orchestrate NegotiaToR's design to be simple yet effective, especially the scheduling algorithm. We believe this is essential when it comes to industrial deployment. 
One may wonder whether the minimalist algorithm is sufficient and whether a slightly more complexity can significantly improve NegotiaToR's performance.
To demonstrate the rationality of our design choices, we investigate several potential variants of NegotiaToR \textbf{(\S\ref{sec:algorithm_design_discussion})}, which are meant to trade off complexity for possible performance gains. This exploration reveals that our minimalist design is sufficient, and additional complexity may not proportionately enhance performance as expected.

\subsection{On-demand distributed scheduling}
\label{sec:negotiator-matching}

We designed a simple matching algorithm, NegotiaToR Matching, for scalable on-demand scheduling, which requires only three steps and no iteration. 
We first show the design of the algorithm itself before going into the big picture of how the algorithm runs on NegotiaToR's fabric and how NegotiaToR sends traffic according to locally derived scheduling results.

\subsubsection{NegotiaToR Matching algorithm}
\label{sub-sec:negotiator-matching-algorithm}

The algorithm runs on ToRs. From real-time binary traffic demand information, ToRs distributedly generate non-conflicting matches for all uplink ports\footnote{Unless specified, we refer to ToR's \textit{uplink port} as \textit{port} in this paper.} so that ToR pairs with traffic demands will get connected. No traffic forecasting is needed. Each ToR is only responsible for traffic flows in and out of it, thus distributing the computational complexity. 
The scheduling process for one epoch is composed of three steps: REQUEST, GRANT, and ACCEPT. These steps distributedly assign links to source-destination port pairs, resolving conflicts at the destinations' and sources' ports, thereby ensuring collision-free data transmission for all ports.

We present the pseudo-code of NegotiaToR Matching in Algorithm \ref{alg-negotiator-matching}. An illustrative example of its workflow is given in Figure \ref{fig:negotiator_matching_algorithm}. The example network consists of eight 4-port ToRs, where the parallel network topology connects them with four 8-port AWGRs, and the thin-clos topology with eight 2-port AWGRs.
Only four ToRs are shown for simplicity.
Note that the algorithm can be used in various flat topologies, including both the parallel network and the thin-clos topology presented in Figure \ref{fig:negotiator_topology} once the GRANT step is modified according to the topology's connection capabilities.

\vspace{6pt}

\noindent\textbf{ToR-level REQUEST}. Each ToR maintains per-destination FIFOs for all ToRs, and data are pushed into the corresponding queue first before heading for their destinations. By checking if the queue has pending data or not, ToRs thus know local traffic demands and notify the destinations by sending requests to them. The requests are ToR-level and are not bound to any specific port. The requests are binary and contain no size or flow-level information, and thus all requests indicate an equal need for link resources. This simplification facilitates subsequent rapid distributed calculation while being sufficient for DCN performance, as we will see later.

\vspace{6pt}

\noindent\textbf{Port-level GRANT}. Now, each destination ToR is aware of the requests from multiple ToRs. To avoid collisions at the destinations' side, the ToR will pick one request for each port in turn. To this end, inspired by RRM \cite{mckeown1995scheduling} used in the scope of crossbar packet switch port scheduling, the destination ToR employs round-robin rings to allocate ports to these requests. The position of the ring's pointer denotes the highest priority source, with priority diminishing in a clockwise direction. After granting one source, the pointer is incremented to the next source of the round-robin schedule. This way, we prioritize the source ToR that's least recently granted, effectively ensuring fairness and avoiding starvation in port allocation.

Depending on the connection capabilities the topology provides, the implementation of this ring differs.
As shown in Figure \ref{fig:the_grant_parallel_network}, there is only one GRANT ring per ToR in the parallel network topology since one port can receive data from any other ToRs. Sharing scheduling states among ports of the ToR thus helps improve scheduling fairness further.
In contrast, in the thin-clos topology, each destination ToR has multiple smaller GRANT rings like in Figure \ref{fig:the_grant_thin_clos}, one for each port. This is because one port in thin-clos can only receive data from a subset of source ToRs \cite{6552916, 6634216}. As a result, grants for a specific port can only be chosen from a subset of requests indicated by the ring.

\vspace{6pt}

\noindent\textbf{Port-level ACCEPT}. After destinations send the grants back, there are possibilities that one port gets multiple grants from different destinations. To resolve this conflict and ensure that only one destination is assigned to one port, the source ToR accepts one grant for each port from received port-level grants, using a per-port ACCEPT round-robin ring for fairness.

\vspace{6pt}

After these three steps, all ToRs are aware of a set of non-conflicting matches indicated by ACCEPT that's derived locally. 
NegotiaToR Matching enables scalable on-demand coordination among ToRs in a distributed manner. It has a comparable scheduling complexity to traffic-oblivious solutions \cite{rotornet, shoal, sirius}, which require congestion control to avoid buffer overflow at intermediate ToRs, similar to NegotiaToR Matching's minimalist request-grant mechanism.
Following the derived scheduling, NegotiaToR can achieve collision-free data transmission through bufferless links.

\subsubsection{Matching efficiency}
\label{sub-sec:matching-efficiency}

We demonstrate NegotiaToR Matching's efficiency in theory with a simplified model, showing that the algorithm has a matching efficiency of 63\% even under high loads where conflicts are frequent.

\newcommand{\FuncCall}[2]{\texttt{\bfseries #1(#2)}}
\SetKwProg{Function}{function}{}{}
\SetKwFunction{REQUEST}{REQUEST}
\SetKwFunction{GRANT}{GRANT}
\SetKwFunction{ACCEPT}{ACCEPT}
\SetAlgoNoLine

\begin{algorithm}[t]
    \caption{NegotiaToR Matching Algorithm}
    \label{alg-negotiator-matching}
    
    \Function{\REQUEST{\textnormal{per-destination queues}}}{
        \ForEach{$queue_i$ \textnormal{\textbf{in} per-destination queues}}{
            \CommentSty{\footnotesize // $queue_i$'s destination is $ToR_i$}

            \If{$queue_i.{has\_pending\_data}$}{
                send a request to $ToR_i$;
            }
        }
    }

    \BlankLine

    \Function{\GRANT{\textnormal{received requests}}}{

        \CommentSty{\footnotesize // allocate ports with round-robin rings}

        $randomly\ initialize\ rings$;

        \ForEach{$port_i$}{ 

            \CommentSty{\footnotesize // get the highest-priority request with per-ToR or per-port GRANT ring}

            $request_x \gets ring_{grant}.get\_request$;

            $port_i.grant \gets request_x$; 

            $ring_{grant}.pointer\_update$;

            send $port_i.grant$ back to $ToR_x$;
        }

    }

    \BlankLine

    \Function{\ACCEPT{\textnormal{received grants}}}{
        \CommentSty{\footnotesize // accept grants with round-robin rings}

        $randomly\ initialize\ rings$;

        \ForEach{$port_i$}{

            \CommentSty{\footnotesize // get the highest-priority grant with per-port ACCEPT ring}

            $grant_x \gets ring_{accept}.get\_grant$;

            $port_i.accept \gets grant_x$; 

            $ring_{accept}.pointer\_update$;

            send traffic to $ToR_x$ through $port_i$;
        }
    }

\end{algorithm}

Consider the following scenario, where we assume $n$ ToRs ($n > 1$) are sending traffic to each other, each having $m$ uplink ports. ToRs are connected by the \textbf{parallel network topology} in Figure \ref{fig:negotiator_topology_big_switch}.
In this model, grants and accepts are randomly and uniformly given. 
In the GRANT step, on average, a source ToR receives grants from $n$ ToRs, and a destination ToR allocates $\frac{m}{n}$ ports per request.
The chance that one port is granted by one destination ToR is thus $\frac{m}{n}/m=\frac{1}{n}$.
Focusing on a specific port ($port_0$) at the source side, the probability of $port_0$ being included in a specific grant ($grant_0$) is thus $\frac{1}{n}$. 
Let $X$ be the number of competing grants for $port_0$, excluding $grant_0$.
$X$ thus follows a binomial distribution $X\sim B(n-1, \frac{1}{n})$. 
The acceptance probability of $grant_0$ by $port_0$ is $Y = \frac{1}{X+1}$. 
The expected value of $Y$ is thus $E(\frac{1}{X+1})=\sum_{k=0}^{n-1} \frac{1}{k+1} P(X=k)$, which is $1-(1-\frac{1}{n})^{n}$.

As $n$ increases, $E[Y]$ monotonically decreases and approaches $1-\frac{1}{e}$. This indicates that when the competition is intense, $grant_0$ still has a 63\% chance to be accepted, and thus $port_0$ at the destination side will get matched. Otherwise, $port_0$ at the destination side will be wasted due to the destination's port being reserved but not accepted by the source ToR. When the same set of ToRs are connected by the \textbf{thin-clos topology}, the matching efficiency itself is typically higher. This is because the number of competing grants for $port_0$ will be smaller due to limited connectivity offered by the topology, leading to a larger value of $E[Y]$.

This conclusion applies symmetrically to all ports, presenting the matching efficiency of NegotiaToR Matching, including at scale. We compare the theoretical results with the simulation results in Appendix \ref{appendix-negotiator-matching}, validating their consistency.

\vspace{6pt}

NegotiaToR Matching is carefully orchestrated to keep simple while fitting in the DCN scenario, like long RTT and high traffic dynamics.
Later in \S\ref{sec:algorithm_design_discussion}, we delve into the rationality of our design choices of NegotiaToR Matching, i.e., the minimalist approach like binary requests and no iteration. In \S\ref{sec:evaluation}, we give comprehensive evaluation results conducted on both the parallel network and thin-clos topology, showing that NegotiaToR Matching achieves high performance in both goodput and FCT across these topologies.

\subsection{Network stack design of control and data planes}
\label{sec:negotiator_network_stack}

There are two expectations for NegotiaToR's network stack. First, it should provide high-frequency connection opportunities among all ToRs to ensure timely exchanges of NegotiaToR Matching scheduling messages. Second, it should provide most of the connection opportunities for the source-destination pairs with traffic demand to ensure high network goodput.
To this end, as we presented in Figure \ref{fig:negotiator_epoch}, each NegotiaToR epoch is split into two phases, the predefined phase and the scheduled phase.

\subsubsection{In-band pipelined scheduling}
\label{sec:inband_control}

The first phase of each epoch frequently reconfigures its connections based on predefined matches and provides all-to-all connectivity. 
Due to the reconfiguration delay introduced by hardware and time synchronization error, each reconfiguration will take a while to finish, 
during which no bits can be sent. For this reason, NegotiaToR inserts guardbands between two timeslots in the predefined phase. 
Note that there is no reconfiguration in the scheduled phase, limiting the impact of guardbands on goodput.

\begin{figure}[t]
    \centering
        \centering
        \includegraphics[width=0.42\textwidth]{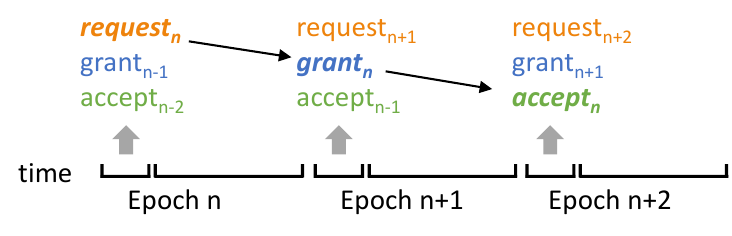}
    \caption{Scheduling for different epochs is performed in a pipelined manner.}
    \label{fig:negotiator_pipelined_exchange}
    \vspace{-5pt}
\end{figure}

The first phase comprises several short timeslots of equal length, and all source-destination pairs will get connected once. 
At the beginning of each timeslot, ToRs reconfigure the wavelengths of their tunable lasers simultaneously according to predefined round-robin matches, establishing all-to-all connections for scheduling message exchange. 
To provide one round of all-to-all connection among $N$ $S$-port ToRs, when forming a parallel network topology, it takes $\lceil{\frac{N-1}{S}}\rceil$ timeslots with the use of $S$ AWGRs, each having $N$ ports. When forming a thin-clos topology, it takes $W$ timeslots using $\lceil\frac{NS}{W}\rceil$ AWGRs, each having $W\,(W\geq\lceil{\frac{N}{S}}\rceil)$ ports.

Utilizing the all-to-all connectivity, ToRs get to exchange scheduling messages periodically. 
To this end, one approach is to finish one scheduling process and get the final scheduling results inside the same epoch. However, the strict front-back dependencies of each scheduling's REQUEST, GRANT, and ACCEPT steps require three rounds of round-robins to finish one scheduling process. 
Meanwhile, a step can start only after the scheduling messages from the previous step are received, thus introducing one-way delay (half RTT) between two consecutive steps. These factors together lead to an extremely long predefined phase, making the network degenerate into a traffic-oblivious design, thus damaging goodput.

Instead, NegotiaToR distributes one scheduling process's $request$, $grant$, and $accept$ into three epochs. Figure \ref{fig:negotiator_pipelined_exchange} illustrates this pipelined workflow. This way, only one round of all-to-all connections is needed in each epoch, shortening the predefined phase. Since we set the scheduled phase to be long to provide enough on-demand sending opportunities, scheduling messages can reach destinations and get processed before the next epoch\footnote{Note that the one-way delay between ToRs and the processing delay together may be longer than one epoch. In this case, the pipelined scheduling still works, only to expand to more epochs.}. Each scheduling thus takes a minimum of around two epochs, facilitating a low scheduling delay.
After the predefined phase, NegotiaToR can send data according to its non-conflicting scheduling results.

\subsubsection{One-hop data transmission}
The second phase establishes connections based on calculated matches, adapting the network to actual traffic demands. 
ToRs set the wavelengths of tunable lasers according to NegotiaToR Matching's scheduling results calculated in the first phase for data transmission.

In this phase, data in the corresponding per-destination queue is sent until the epoch ends or the queue empties, satisfying the traffic demands and contributing to a high goodput. 
Regarding latency, because of NegotiaToR's conflict-free scheduling, even though we aimed to design a low-latency algorithm and employ no iteration, a scheduling delay of at least two epochs is inevitable for a previously empty source-destination pair that has newly arrived data.
Moreover, extra waiting delay will be introduced if this request is not granted. For elephant flows, this is acceptable. However, for mice flows that require low latency, this will essentially damage their FCT. FCT-oriented optimizations are needed.

\subsection{Incast-optimized scheduling delay bypass}
\label{sec:bypass_scheduling_delay}

In DCNs, other than bandwidth-sensitive elephant flows, there are also latency-sensitive mice flows, which occupy a small portion of the total traffic volume but represent a large portion when measured by the number of flows \cite{homa}. These mice flows often appear as incasts where multiple mice flows arrive simultaneously in burst and compete for the limited scheduling opportunities. They require small FCT for good application performance, which is hard to achieve considering scheduling delays. The performance degradation worsens when incast happens.

\subsubsection{Data piggybacking in the predefined phase}
\label{sec-piggybacking}
\begin{figure}[t]
    \centering
        \centering
        \includegraphics[width=0.43\textwidth]{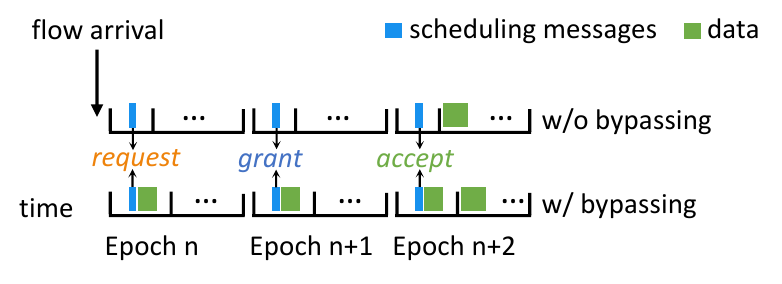}
    \caption{Bypassing scheduling delay with unscheduled transmission in the predefined phase.}
    \label{fig:negotiator_bypass_on_and_off}
    \vspace{-5pt}
\end{figure}

To optimize mice flows' FCT especially under incast, NegotiaToR offers unscheduled transmission opportunities that bypass the scheduling delay. 
In the first phase, NegotiaToR piggybacks one small data packet along with scheduling messages utilizing the predefined all-to-all connectivity, as we showed in Figure \ref{fig:negotiator_bypass_on_and_off}. 
Therefore, each ToR pair is guaranteed the transmission of at least one packet in every epoch, regardless of scheduling results.

We limit the size of piggybacked data to be small so that the predefined phase can still be short, ensuring that the majority of connections in one epoch are established in the scheduled phase based on actual traffic demands for high goodput. This is sufficient for mice flows considering their small size.
Mice flows thus can be sent promptly via one-hop paths without suffering from scheduling delays, even under incasts, contributing to their small FCTs.

After deploying such a piggybacking mechanism, we apply a slight adjustment to NegotiaToR Matching---requests can only be sent when the queued data in per-destination queues exceeds three piggybacked packets instead of zero.
This is because, during the scheduling delay, three packets are guaranteed to be sent in predefined phases of three consecutive epochs, regardless of the scheduling result. Raising the request threshold thus can avoid providing connections to source-destination pairs with no traffic demand, reducing bandwidth waste.

\subsubsection{Priority queue for mice flow prioritization}
\label{sec:mice_flow_prioritization}

Note that elephant flows in source ToRs, whose FCT is not our concern, may block mice flow data and occupy precious unscheduled transmission opportunities. 
In order to maximize the benefits of the piggybacked packet and mitigate such head-of-line blocking, NegotiaToR deploys mice flow priority mechanism in ToRs. We adopted an existing solution---the information-agnostic mice flow priority mechanism, PIAS \cite{pias}. By utilizing multi-level feedback queues, it can send mice flow data first without requiring prior knowledge of flow sizes, maintaining the practicality of NegotiaToR's design. This facilitates mice flows' timely transmission in both the predefined and scheduled phases.

With data piggybacking in the predefined phase and mice flow priority queues in ToRs, mice flow FCT is reduced, most of which is even below the scheduling delay. We evaluate the effectiveness of these two design elements through microbenchmarks, detailed in \S\ref{sec:evaluation_micro}. So far, we have presented the design of NegotiaToR's network stack. 
We summarize the role of the two phases in Table \ref{fig:data_and_control_plane}.

\subsection{The rationality of NegotiaToR's design choices}
\label{sec:algorithm_design_discussion}

When designing the on-demand scheduling algorithm, we are committed to ensuring performance while reducing complexity for deployment practicality. This leads to our design choices towards a minimalist design, like binary requests and no iteration.
This subsection delves into a common question: \textit{could a more complex design significantly improve NegotiaToR's performance?} 
We examine this possibility, analyzing the reasoning behind our minimalist design, and demonstrate that extra complexity does not necessarily translate into proportionate performance gains. Note that we do not claim that our design is the ultimate solution. It remains an open problem to optimally balance complexity with performance, inviting further investigation and potential enhancements.

\vspace{6pt}

\noindent{\textbf{No iteration.}} 
NegotiaToR adopts a non-iterative design, distinguishing it from traditional iterative matching algorithms like PIM \cite{DBLP:journals/tocs/AndersonOST93}, RRM \cite{mckeown1995scheduling} and iSLIP \cite{islip}. 
Iteration can potentially improve matching efficiency and goodput, but also increases complexity and scheduling delays, especially in DCNs where the RTT between ToRs is long when put into the context of nanoseconds reconfiguration delay. With longer scheduling delays, the scheduling result is more likely to be outdated, since previous epochs may have already sent all the data during the long scheduling process, leaving the scheduled links empty, wasting bandwidth and thus adversely affecting performance. 

We design an iterative version of NegotiaToR Matching, and investigate the impact of iteration through simulations. Details can be found in \textbf{Appendix \ref{sec:iterative-negotiator-matching}}. 
Simulation results confirm that using iteration is not a good idea to improve goodput for NegotiaToR. The iterative approach yields minimal or even negative improvements in goodput due to the outdated scheduling results, and consistently exhibits worse FCT due to longer scheduling delays. 
In contrast, simply using a 2$\times$ link rate speedup (i.e., the bandwidth ratio between uplinks and downlinks for each ToR is 2:1) upon the original non-iterative scheduling, which is a common practice in optical switching, achieves high goodput and superior FCT, making the iteration process unnecessary. As a result, we use $2\times$ speedup in the following exploration.

\vspace{6pt}

\begin{table}[]
    \centering
    \footnotesize
    \begin{tabular}{|c|c|c|}
    \hline
     &
      Predefined Phase &
      Scheduled Phase \\ \hline
    \begin{tabular}[c]{@{}c@{}}In-band\\ Control Plane\end{tabular} &
      \begin{tabular}[c]{@{}c@{}}on-demand\\ distributed scheduling\end{tabular} &
      / \\ \hline
    \begin{tabular}[c]{@{}c@{}}One-hop\\ Data Transmission\end{tabular} &
      \begin{tabular}[c]{@{}c@{}}unscheduled transmission\\ (bypass scheduling delay)\end{tabular} &
      \begin{tabular}[c]{@{}c@{}}scheduled\\ transmission\end{tabular} \\ \hline
    \end{tabular}
    \vspace{0.1in}
    \caption{Two phases' role as the control plane and data plane in one NegotiaToR epoch.}
    \label{fig:data_and_control_plane}
    \vspace{-5pt}
\end{table}

\noindent{\textbf{No data relay.}} 
NegotiaToR chooses direct one-hop paths for all data, instead of traffic-oblivious relay \cite{rotornet, shoal, sirius}. There exists a third design choice, traffic-aware selective relay, which only enables data relay for elephant flows under light loads \cite{trod}. This avoids goodput damages of the traffic-oblivious one under heavy loads as well as mice flow FCT damages caused by more hops. By fully utilizing the empty links, it may potentially improve goodput for the thin-clos topology (Figure \ref{fig:negotiator_topology_thin_clos}) where the connectivity is limited.

To figure out if the performance gain is worth the added complexity, like intermediate ToR selection and congestion control for the relayed traffic, we carefully design a traffic-aware selective relay algorithm and conduct simulations on the thin-clos topology. Details can be found in \textbf{Appendix \ref{sec:traffic-aware-intermediation}}. 
Results show minor or no goodput gain, because NegotiaToR's goodput is already good at light loads, while the data relay barely helps at heavy loads. 
Considering the complexity of implementing data relay, we conclude that relay is not a good fit for NegotiaToR.

\vspace{6pt}

\noindent{\textbf{Binary requests.}} 
Another potential improvement is to include more information in the requests indicating priorities, instead of using binary demand information and simply prioritizing the least recently allocated pair with round-robin rings. We explore two approaches: one goodput-oriented approach that prioritizes the pairs with more pending data to improve link utilization, and one FCT-oriented approach that prioritizes the pairs with longer head-of-line packet waiting delays to reduce tail FCT. 

To study if the benefits outweigh the additional complexities of size measurement, delay logging, and sorting, we evaluate the two approaches through simulations. We attach the details in \textbf{Appendix \ref{sec:informative-requests}}. Results indicate that for both approaches, the performance improvements are relatively modest compared with the added complexity. We thus conclude that binary requests are simple and effective enough for NegotiaToR.

\vspace{6pt}

\noindent{\textbf{Stateless scheduling.}} 
In our design, ToRs send requests solely according to real-time demands, without tracking ongoing requests from previous epochs. This can lead to over-scheduling the same source-destination pair, potentially resulting in under-utilized links and hurting goodput when all data has already been sent at the time of acceptance. 
Incorporating a stateful traffic matrix can mitigate this, but will introduce the complexity of maintaining the states, which reduces the robustness of the system especially under failures.
Moreover, for lightly loaded scenarios, the impact of duplicate requests is negligible since the wasted links are not needed otherwise. For heavily loaded scenarios, since the data arrive at the sources continuously, whenever scheduled, the sources can always fully utilize the links, which can even decrease FCTs. The impact of stateless scheduling is thus minor.

To validate this, we designed and simulated a stateful version of NegotiaToR Matching, as detailed in \textbf{Appendix \ref{sec:stateful-scheduling}}. The results confirm our analysis, showing a negligible difference between stateful and stateless scheduling. 
This justifies our design choice.

\vspace{6pt}

For now, we have found that the possibilities discussed above do not yield significant performance gains, and the added complexity is not worth the cost. Therefore, NegotiaToR converges to its present form, which is orchestrated towards a minimalist design. 
We acknowledge that our exploration has not covered the entire design spectrum. 
As DCN applications continue to develop, we believe there are opportunities for further optimizing NegotiaToR. We leave this as future work.

\subsection{Practical considerations}
\label{sec:practical_considerations}

When deploying NegotiaToR in real-world DCNs, it's crucial to address practical concerns such as link failure handling. This section delves into how NegotiaToR manages these practical problems.

\subsubsection{Fault tolerance}
\label{sec:fault_tolerence}

NegotiaToR incorporates a mechanism for handling link failures. In the predefined phase, even when there is no scheduling message to send, we intentionally let each ToR send a dummy message to distinguish between link failures and lack of traffic.
The loss of messages in the first phase indicates possible link failures. 
ToRs also employ this message to give feedback on whether they have received any bits in the reverse direction, along with port information if applicable.

This way, a ToR can effectively determine whether one port's egress or ingress link has failed\footnote{We detect faults of egress and ingress separately to prevent overreaction and simplify maintenance tasks.}. 
In the predefined phase, each port in one ToR is expected to receive bits from multiple source ToRs. An ingress link failure may have happened if a ToR consistently fails to receive bits through a particular port.
Similarly, on the sender side, repeated feedbacks of undelivered data originating from a specific port indicate an egress link fault.
Upon detecting a link failure, ToRs will alert the maintainers and broadcast the detected failures to update their scheduling rules. 
This involves excluding the affected egress links and ingress links from data transmission. Once the faulty links are repaired, the transmission of scheduling messages can resume, and the corresponding links can be included in the scheduling again.
ToRs can resume data transmission through them.
As for the recovery of lost packets during handling the failure, NegotiaToR relies on upper-layer protocols, like TCP, for retransmission.

To further reduce link waste, for the parallel network topology, we also periodically change the round-robin rule in the predefined phase. 
This approach allows a pair of ToRs to exchange scheduling messages through multiple port-to-port links, instead of a specific one. Consequently, if one link experiences a failure, the exchange of scheduling messages with the corresponding subset of ToRs can still proceed through other links in subsequent epochs, guaranteeing all-to-all connectivity in the predefined phase. This ensures that the affected ToR pairs can still transmit data through the remaining functional links in the scheduled phase.
For the connection-limited thin-clos topology where one source-destination pair can only communicate through identical ports, the same goal can be achieved by relaying scheduling messages and data to an intermediate ToR through ports that function normally.
We test this fault detection and recovery mechanism on the parallel network topology in our evaluation ({\S\ref{sec:evaluation_results}}).

\subsubsection{Scalable scheduling logic implementation}
\label{sec:scheduling_logic_implementation}

Since NegotiaToR Matching draws inspiration from RRM \cite{mckeown1995scheduling} that focuses on port matching inside crossbar packet switches, which is a well-mined area, it can benefit from applying proven implementation strategies in this field regarding hardware implementation, especially for the GRANT and ACCEPT rings.
Switch port matching algorithms \cite{mckeown1995scheduling, islip} use programmable priority encoders \cite{748793} to construct round-robin arbiters to function as priority rings.
One iteration of matching can be done in several clock cycles for a switch of hundreds of ports \cite{1205968}. For the GRANT and ACCEPT rings in NegotiaToR Matching, the implementation is similar, but with a reduced complexity as the transition from the individual-switch scale to the DCN scale---this transition decentralizes the scheduling logic of ports from one single chip to multiple ToRs, and also extends the scheduling time budget compared to that inside switches. These factors collectively underscore the high practicality of implementing NegotiaToR Matching logic at scale.

\subsubsection{End-to-end reconfiguration delay}
\label{sec:reconfiguration_delay}

NegotiaToR employs frequent network reconfigurations to serve the dynamic traffic demands among ToRs, thus preferring a shorter end-to-end reconfiguration delay to reduce guardband overhead. 
This calls for fast wavelength tuning hardware, fast clock and data recovery (CDR), and precise time synchronization so that a smaller guardband is sufficient to absorb the reconfiguration delay.
Recent advancements in tunable lasers and CDR mechanisms have lowered the tuning delay to 10s of nanoseconds \cite{8976124, sirius, 9333367}. In particular, \cite{sirius} employs disaggregated tunable lasers \cite{9125762}, along with amplitude caching \cite{9489945} and clock phase caching \cite{8535333, 8535334}\footnote{For NegotiaToR, ToRs read the $accept$ packets to access the connections to be established, so that they can directly use the cached parameters for fast CDR.}, and can limit tuning and CDR delays to under 10 nanoseconds.

As for time synchronization, recent studies have lowered synchronization errors to a few 10s of nanoseconds for a conventional packet-switched DCN \cite{211255}. 
For architectures like NegotiaToR where the AWGR is passive and no retiming or queueing latency jitter is involved, synchronization error can be further reduced.
For example, \cite{sirius} leverages round-robin all-to-all connections for synchronization, achieving errors within picoseconds. 
A primary clock is chosen, and ToRs periodically synchronize their local clocks and times with this primary.
Similar mechanisms can be applied to NegotiaToR using the round-robin connections in the predefined phase. 
After synchronization, the clocks will slowly drift \cite{211255}, and a guardband of several nanoseconds is adequate to absorb the drift till the next synchronization in the next predefined phase.

Together, current technologies can facilitate an end-to-end reconfiguration delay of 10 nanoseconds as demonstrated in \cite{sirius}, leading to our main reconfiguration delay setting in \S\ref{sec:evaluation}. 
Note that even if the guardband is enlarged to 100 nanoseconds, NegotiaToR remains effective with appropriate parameter modifications. We evaluate the influence of different guardbands in \S\ref{sec:evaluation_micro}.

\subsubsection{Epoch length setting considerations}
\label{sec-epoch-length}
We discuss the length settings of the two phases of one NegotiaToR epoch. First of all, we do not want the predefined phase to dominate, as the connections are irrelevant to current traffic demands and will downgrade to pure round-robin, resulting in a huge performance drop.
Conversely, an extremely short predefined phase limits opportunities for scheduling delay bypassing, adversely affecting mice flows. 
Meanwhile, a scheduled phase shorter than one-way delay between ToRs is also undesirable due to increased scheduling delay.
Furthermore, for a low scheduling delay and high scheduling frequency, the epoch length should not be excessively long, which leads to increased FCT and possible link waste due to outdated scheduling results.

To minimize overhead, guardbands before wavelength tunings are desired to account for 10\% or less of all time.
For a low wavelength tuning delay of 10 nanoseconds \cite{sirius}, it's rather easy to reach this goal. 
For longer guardbands, extending the scheduled phase proportionally, which does not involve reconfigurations, also meets this target.
This leads to our default epoch length settings in evaluation ({\S\ref{sec:evaluation_setup}}). We conduct simulations to show the influence of different guardbands in \S\ref{sec:evaluation_micro}.
In \S\ref{sec:evaluation_deep_dive}, parameter sensitivity experiments are also conducted to further understand the impact of epoch length settings on performance.

\subsubsection{Traffic management below ToRs}

Our design focused on the interconnections above ToRs. Here, we discuss possible methods to do traffic control below ToRs.
Traffic from hosts to ToRs is buffered at the ToR before transmission through the optical fabric.
Backpressure-based or credit-based flow control can help to avoid packet drops, which stop corresponding data transmission when the buffer is full. 
On the receiver side, data arriving at destination ToRs are buffered before transmission to hosts. Since we introduce speedup in the optical fabric, and data destined for the same host may arrive synchronously at the ToR through multiple ports, the corresponding queue in the receiver side ToR may accumulate and cause packet drops. 
To cope with this, ToRs should monitor the length of this queue and only allow data transmission when buffer space is enough.
This slight modification thus reduces packet drops.

To ensure reliability, we can run TCP-like protocols over NegotiaToR, which typically involve reordering at the receiver side. 
NegotiaToR's design inherently results in fewer out-of-order packets. 
For one source-destination pair, as long as the packets are moved from the per-destination queue to per-port queues in order at the source ToR (e.g., moving to lower-index port first if there are multiple available ports), and then consumed in the same order at the destination ToR, the order of packets will be preserved.

\section{Evaluation}
\label{sec:evaluation}

We investigate the effectiveness of NegotiaToR's design elements, and evaluate its performance through large-scale simulations using a packet-level simulator, YAPS \cite{phost}. Simulations on the parallel network topology and the thin-clos topology are done, validating the generality of NegotiaToR on flat topologies.

\subsection{Evaluation setup}
\label{sec:evaluation_setup}

\noindent {\textbf{Network setup.}} The network consists of 128 8-port ToRs, meaning each ToR connects to the optical fabric through 8 ports. We connect the same set of ToRs with the parallel network topology (Figure \ref{fig:negotiator_topology_big_switch}) with 8 128-port AWGRs, and the thin-clos topology (Figure \ref{fig:negotiator_topology_thin_clos}) with 64 16-port AWGRs, respectively. One-way propagation delay between ToRs is $2 \mu s$. The hosts under the same ToR have an aggregated bandwidth of 400 Gbps, and we provide a $2 \times$ link rate speedup to ToR uplink ports (i.e., 100 Gbps per port, and $8 \times 100$ Gbps total), as discussed in \S\ref{sec:algorithm_design_discussion}.  Focusing on ToR interconnections, we consider ToRs as endpoints. FCT and goodput measurements are taken from the ToRs' perspective, marking the start and end of flows at the ToRs. 
Unless specified, data piggybacking (PB) in the predefined phase for scheduling delay bypass and priority queues (PQ) at sources for mice flow prioritization (\S\ref{sec:bypass_scheduling_delay}) are both enabled. For PQ, we set three priorities: the first $1KB$ of flow data will be sent first, then the following $9KB$, and then the rest of the bits.

\vspace{6pt}

\noindent{\textbf{Baseline.}} 
We compare the performance of NegotiaToR with traffic-oblivious proposals on the same scale network (i.e., 128 8-port ToRs), 
following Sirius \cite{sirius} to implement the state-of-the-art benchmark on the same simulator. 
Note that while NegotiaToR can be customized for different flat topologies, Sirius is specifically designed for the thin-clos topology. Its relay-enabled round-robin scheduling cannot utilize the sufficient connectivity of the parallel networks, resulting in identical performance on both topologies. \textbf{Therefore, for brevity, we only show the results on the thin-clos topology for the traffic-oblivious scheme}. $2\times$ speedup and priority queue for mice flow prioritization are also enabled. 
Since the multi-level-feedback-queue based prioritization \cite{pias} does not apply to data at intermediate nodes, we only enable it at sources.

\vspace{6pt}

\noindent {\textbf{Epoch settings.}} Based on \cite{sirius}, the guardband for reconfiguration in NegotiaToR is also set to $10 ns$ unless specified.
By default, in the predefined phase, each timeslot takes $60 ns$, comprising a $10 ns$ guardband and $50 ns$ for transmitting NegotiaToR Matching scheduling messages along with data packet header ($30 B$ each, including $request$, $grant$, and $accept$) plus a data payload ($595 B$).
During the scheduled phase, no reconfiguration or guardband is required, with each timeslot lasting $90 ns$ for sending one data packet (including a $10 B$ header). We set the length of the scheduled phase to $30$ timeslots to balance goodput and FCT unless specified.
Consequently, for both topologies, the predefined phase takes $16 * 60 ns = 0.96 \mu s$, and the scheduled phase takes $30 * 90 ns = 2.7 \mu s$. This leads to an epoch size of $3.66 \mu s$, where the guardbands account for $4.37 \%$. 
We assume that the one-way delay between ToRs ($2 \mu s$) and the calculation delay together is lower than one epoch, leading to a scheduling delay of around two epochs.

\vspace{6pt}

\noindent {\textbf{Workload characteristics.}} Unless specified, we generate the workload after published DCN traces collected from Meta's Hadoop clusters \cite{DBLP:conf/sigcomm/RoyZBPS15}. The trace is highly tailed, where 60\% of the flows are less than $1KB$, while more than 80\% of the bits are from elephant flows larger than $100KB$. All the flows arrive based on a Poisson process, with sources and destinations chosen uniformly at random. We define the network load as $L = \frac{F}{R\cdot N \cdot \tau}$. $F$ is the mean flow size, $R$ is the per-ToR bandwidth, $N$ is the number of ToRs, and $\tau$ is flows' inter-arrival time.
Since we view the network as starting from ToRs instead of hosts, $R$ is the aggregated bandwidth of the hosts under one ToR, which is 400 Gbps. We test the network's goodput and FCT under various loads, ranging from 10\% to 100\%. 
Other than the Hadoop workload, NegotiaToR's performance under various workloads is also evaluated later in \S\ref{sec:evaluation_deep_dive}.

\vspace{6pt}

\noindent {\textbf{Evaluation metric.}} Unless noted, we simulate a real-world duration of $30 ms$, and focus on 99th-percentile mice flow FCT, as well as the average goodput of all ToRs. Flows less than $10KB$ are regarded as mice flows. If not stated otherwise, goodput is normalized to the aggregated bandwidth of the hosts under one ToR (400 Gbps).

\subsection{Microbenchmarks}
\label{sec:evaluation_micro}

We first understand the advantages of NegotiaToR's design elements through microbenchmarks.

\vspace{6pt}

\noindent{\textbf{The effectiveness of NegotiaToR's scheduling delay bypassing designs.}} We conduct ablation studies on our designs for bypassing scheduling delays, i.e., data piggybacking (\S\ref{sec-piggybacking}) and mice flow prioritization (\S\ref{sec:mice_flow_prioritization}). The results are presented in Table \ref{tab:mice_optimization1backup}.
When data piggybacking is disabled, the timeslot in the predefined phase is shortened with only reconfiguration delays and scheduling messages left, and the scheduled phase is enlarged to keep the epoch length the same, accordingly maintaining the reconfiguration overhead ratio the same.
At 100\% load, enabling only data piggybacking itself largely reduces the 99th percentile FCT of mice flows compared to no optimization.
When combined with priority queues, the head-of-line blocking of elephant flows is mitigated, allowing unscheduled piggybacking opportunities to be better utilized by mice flows. 
For both topologies, mice flow is thus further optimized, where the average FCT drops to 1.6 epochs, which is even less than the roughly 2-epoch scheduling delay. 

\begin{figure}
    \centering
    \vspace{-5pt}
        \includegraphics[width=0.45\linewidth]{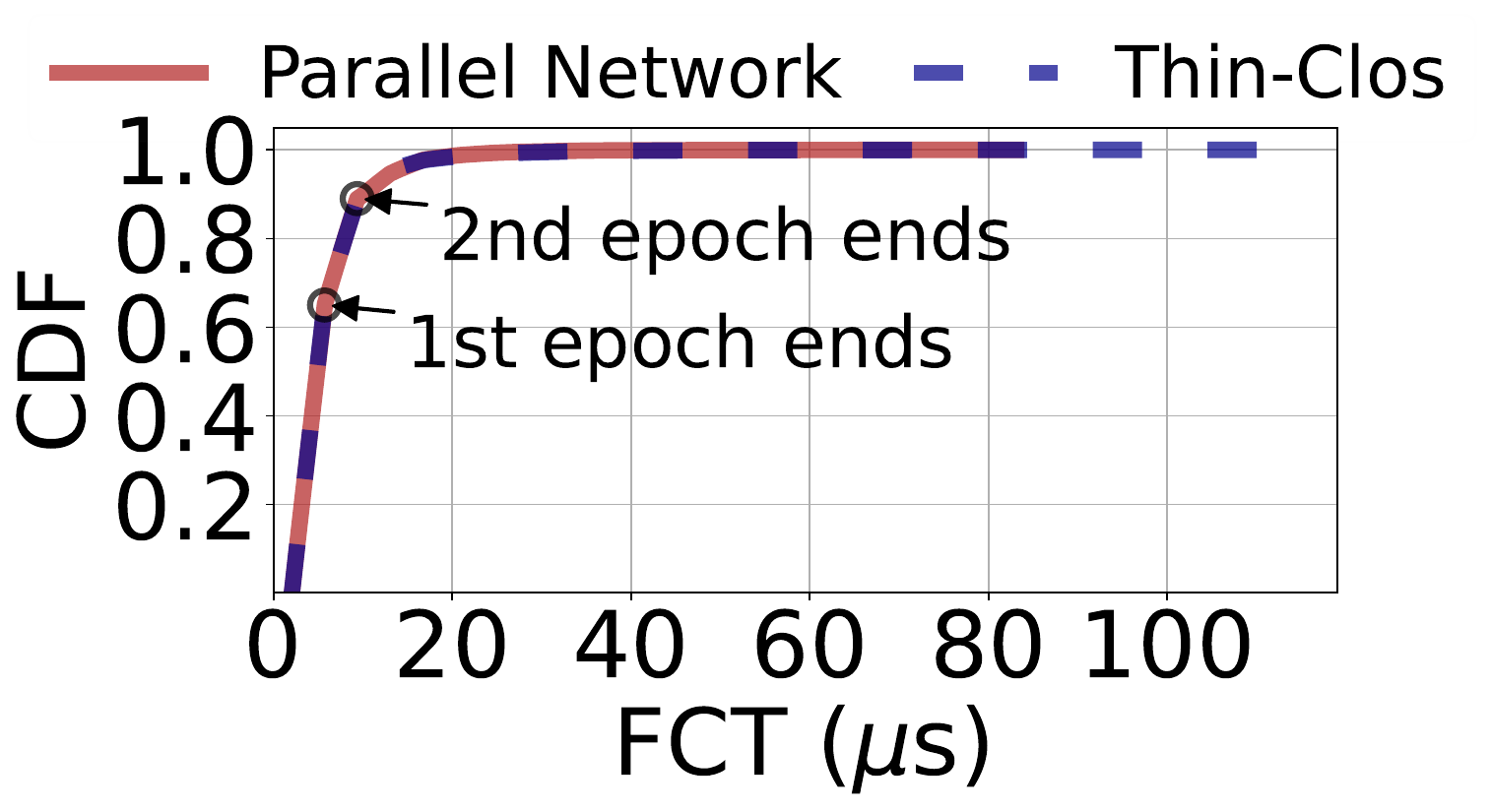}
    \vspace{-7pt}
    \caption{CDF of NegotiaToR's mice flow FCT at 100\% load.}
    \label{fig:mice_optimization_CDF}
    \vspace{-5pt}
\end{figure}

To delve deeper into NegotiaToR's mice flow performance, we show the CDF of mice flow FCT at 100\% load with both PB and PQ enabled in Figure \ref{fig:mice_optimization_CDF}. 
Both topologies provide identical connectivity in the predefined phase, leading to the overlapping of two lines for smaller FCTs. Across two topologies, over 80\% of mice flows successfully bypass the scheduling delay, finishing within 2 epochs (the second turning point). This validates the effectiveness of our designs of bypassing scheduling delays. Mice flow performance thus can be guaranteed even under heavy loads.

\begin{table}[t]
    \centering
    \footnotesize
    \begin{tabular}{|l|c|c|}
    \hline
    \multirow{2}{*}{} & \multicolumn{2}{c|}{Mice Flow FCT in Epochs (99p/Average)} \\
    \cline{2-3}
    & Parallel Network & Thin-Clos \\
    \hline
    - & 732.4/42.1  & 1216.4/75.0 \\
    \hline
    PB & 418.5/19.9  & 847.9/45.3  \\
    \hline
    PQ & 21.0/5.7  &  26.4/5.7 \\
    \hline
    PB and PQ & 6.0/1.6 & 6.5/1.6  \\
    \hline
    \end{tabular}
    
    \vspace{0.1in}
    \caption{NegotiaToR's mice flow FCT at 100\% load, with data piggyback (PB) in the predefined phase and priority queues (PQ) separately enabled and disabled.}
    \label{tab:mice_optimization1backup} 
    \vspace{-10pt}
\end{table}

\begin{figure}[t]
    \centering
    \begin{minipage}[h]{0.4\textwidth}
        \centering
        \includegraphics[width=\linewidth]{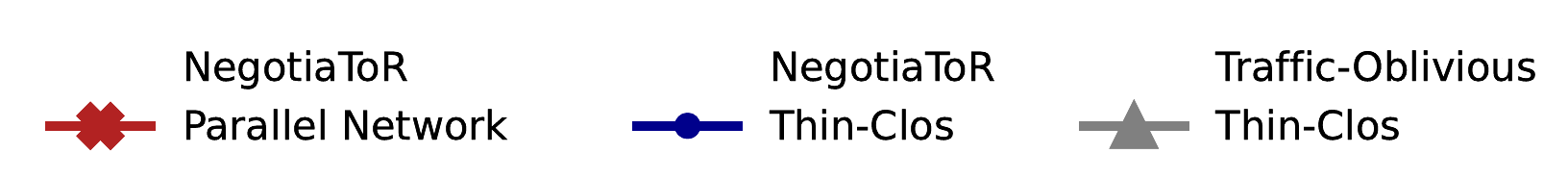}
        \vspace{-0.25in}
    \end{minipage}
    \hfill
    \subfigure[Incast finish time]{
        \begin{minipage}[b]{0.18\textwidth}
        \centering
        \includegraphics[width=\linewidth]{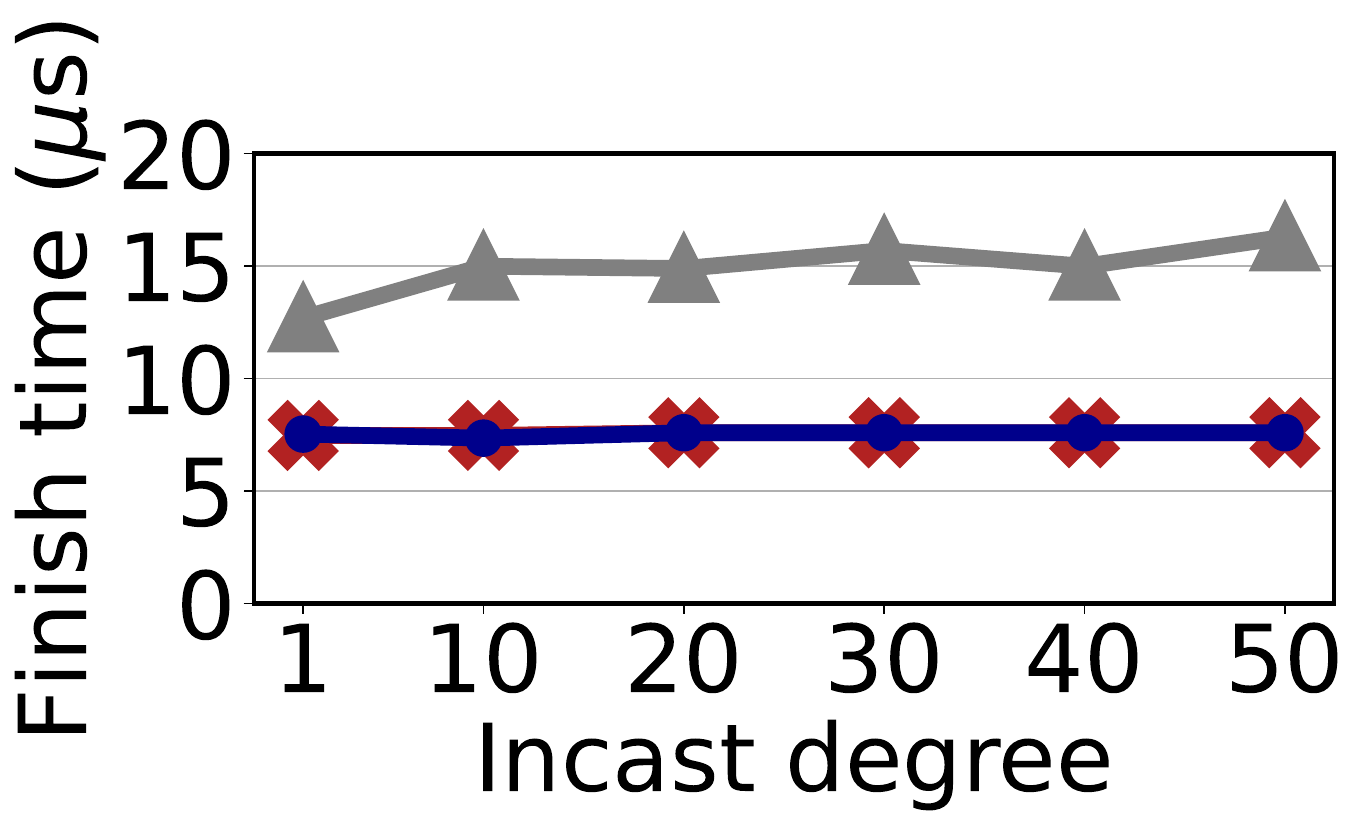}
        \vspace{-0.2in}
        \label{fig:incast_finish_time}
        \end{minipage}
    }
    \hfill
    \subfigure[Average goodput under all-to-all]{
        \begin{minipage}[b]{0.18\textwidth}
        \centering
        \includegraphics[width=\linewidth]{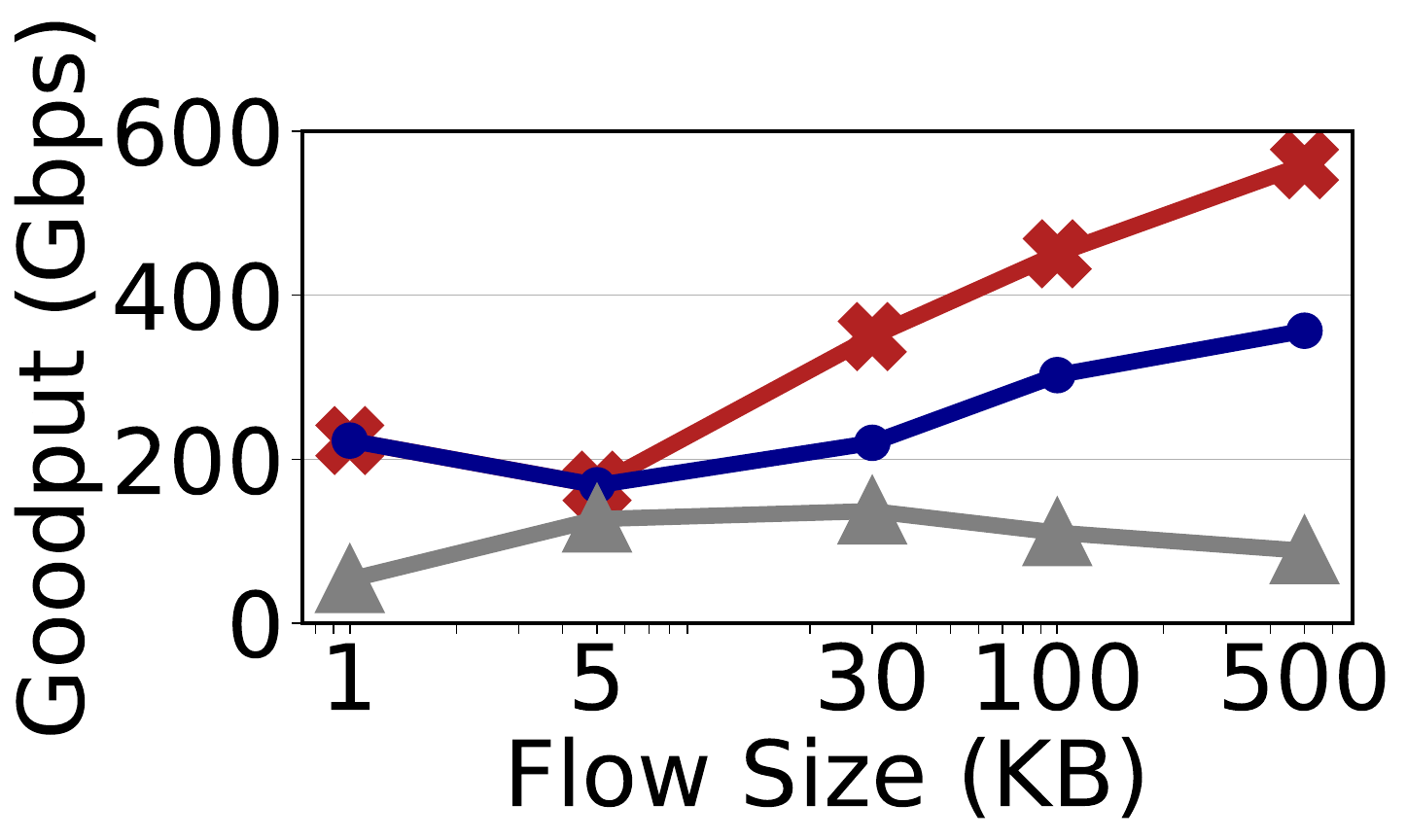}
        \vspace{-0.2in}
        \label{fig:alltoall_goodput}
        \end{minipage}
    }
    \hfill
    \vspace{-7pt}
    \caption{Performance under incast and all-to-all loads.}
    \vspace{-5pt}
    \label{fig:motivation_macro}
\end{figure}

\vspace{6pt}

\noindent{\textbf{Incast finish time.}} To further verify NegotiaToR's incast-optimized scheduling delay bypass design, 
we test incast workloads with varying incast degrees on NegotiaToR and the traffic-oblivious scheme, where a set of ToRs synchronously send one $1 KB$ flow to the same ToR, and the number of source ToRs is the degree. Results are shown in Figure \ref{fig:incast_finish_time}. 
Compared with the traffic-oblivious scheme, NegotiaToR consistently finishes the incast earlier at roughly the same time by piggybacking data in the predefined phase without scheduling, regardless of the incast degree.
Note that NegotiaToR achieves almost the same incast finish time on two topologies. 
Again, this is because both topologies provide identical connectivity in the predefined phase. 
To gain further insights, we observe the goodput at the receivers' side, with results presented in Appendix \ref{sec-appendix3}.

\vspace{6pt}

\noindent{\textbf{All-to-all goodput.}} NegotiaToR Matching can provide sufficient connectivity for traffic in an on-demand manner, even at heavy loads. 
We test all-to-all workloads with varying flow sizes to verify this, where each ToR synchronously sends equal-sized flows to all other ToRs. Average goodput during the transmission is shown in Figure \ref{fig:alltoall_goodput}. For heavier loads, NegotiaToR fully employs the $2\times$ speedup, achieving much higher goodput. Its goodput on the parallel network is better than that on the thin-clos topology because of the latter's limited connectivity: with flows completing, the thin-clos topology experiences underutilization of links, whereas links in the parallel network maintain higher utilization. In contrast, even though with a $2\times$ speedup, the goodput of the traffic-oblivious scheme is limited since the network is flooded by relayed traffic, which competes for receivers' bandwidth, thus damaging goodput, especially at heavy loads.
We also make micro-observations, with detailed results presented in Appendix \ref{sec-appendix3}.

\vspace{6pt}

\begin{figure}
    \centering
  
    \subfigure[Goodput and mice flow FCT on parallel network]{
        \begin{minipage}[b]{0.22\textwidth}
          \centering
          \includegraphics[width=\linewidth]{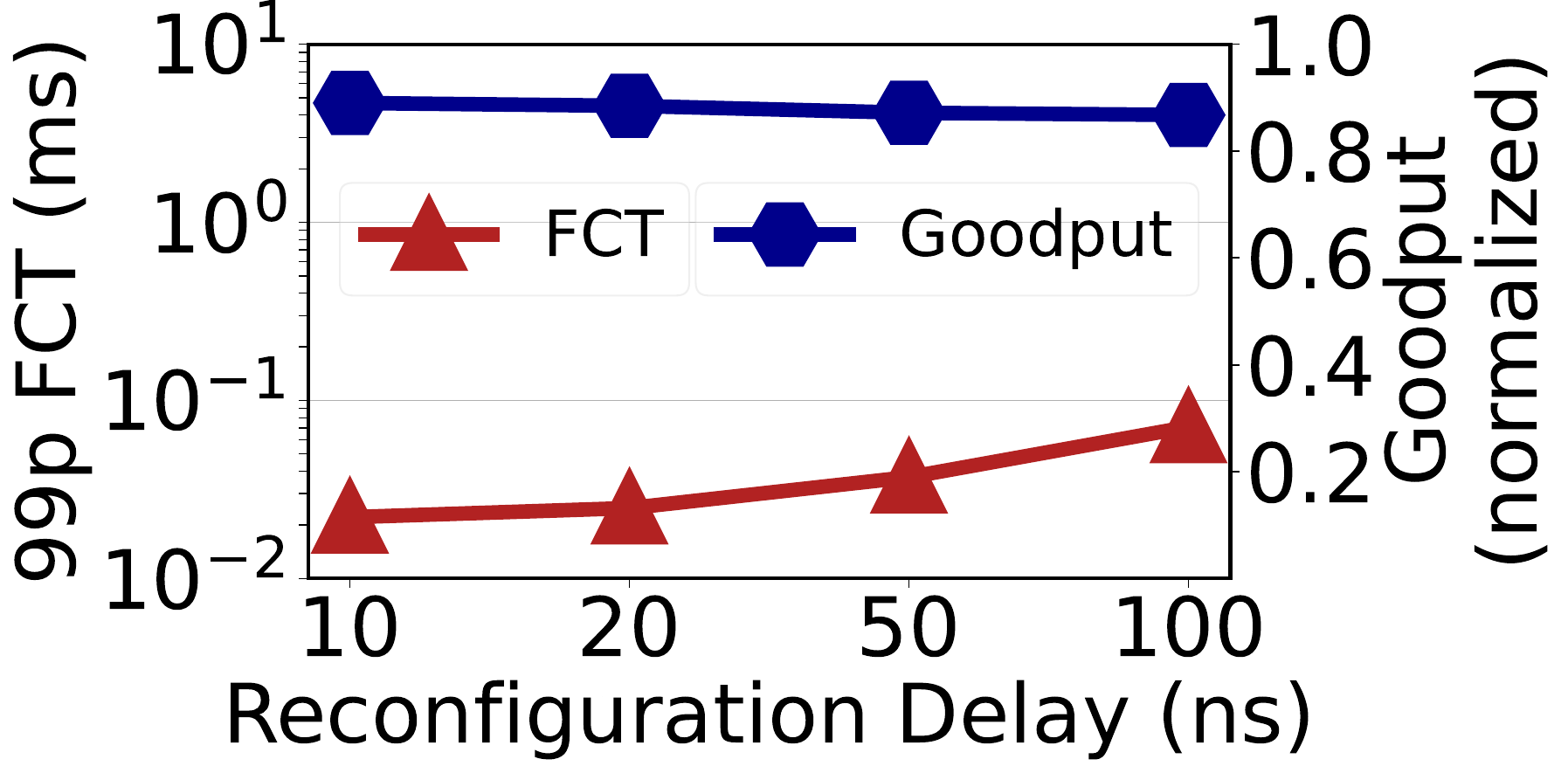}
        \end{minipage}
    }
    \hfill
    \subfigure[Goodput and mice flow FCT on thin-clos]{
        \begin{minipage}[b]{0.22\textwidth}
          \centering
          \includegraphics[width=\linewidth]{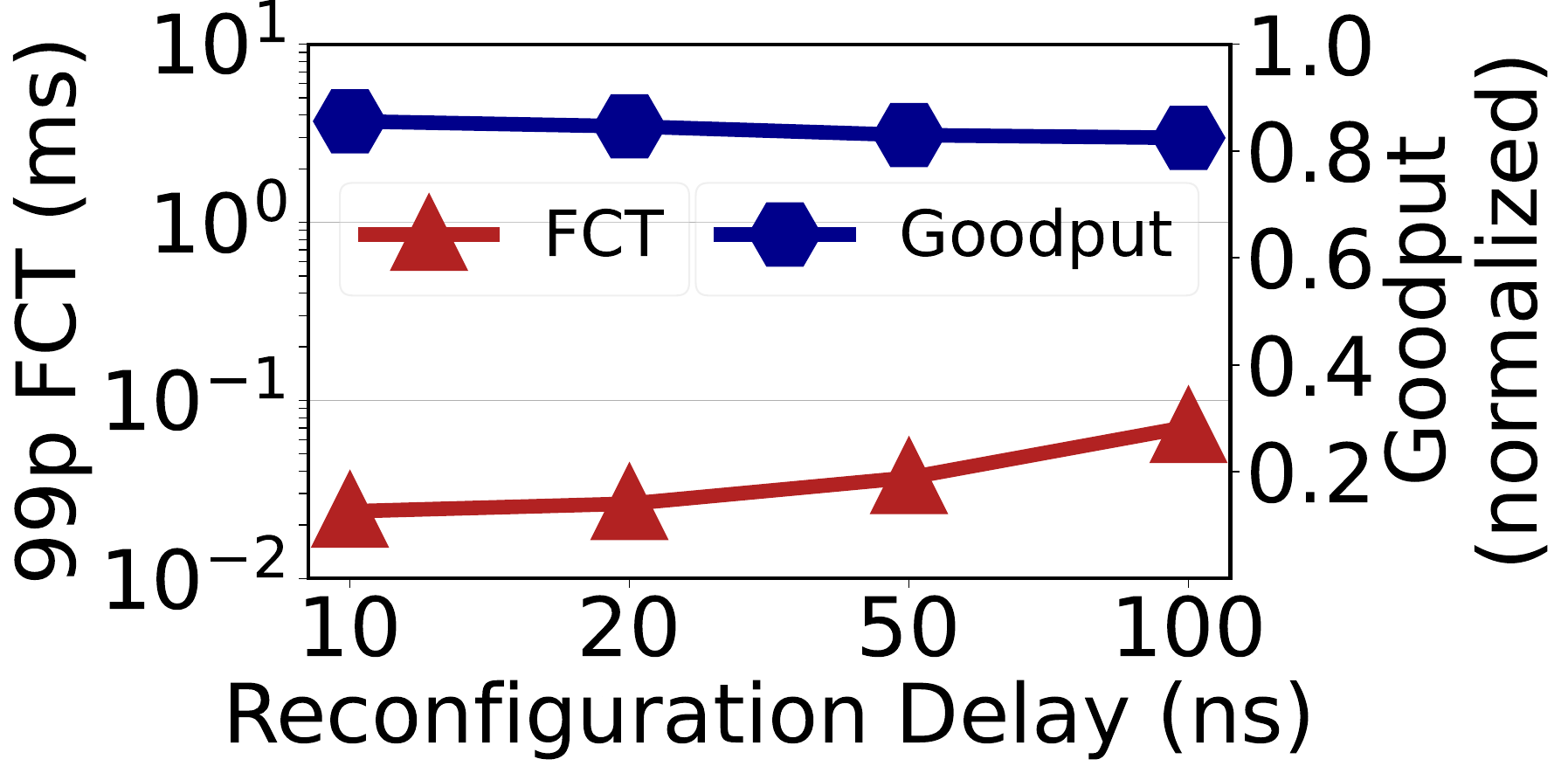}
        \end{minipage}
    }
    \vspace{-5pt}
    \caption{NegotiaToR under various reconfiguration delays at 100\% load.}
    \label{fig:reconfiguration}
    \vspace{-10pt}
\end{figure}

\noindent{\textbf{Adaptability to various reconfiguration delays.}} Despite $10 ns$ end-to-end reconfiguration delays already being realized \cite{sirius}, we also evaluate NegotiaToR's performance under longer reconfiguration delays at 100\% load. The length of the scheduled phase is accordingly adjusted to control the reconfiguration overhead. The findings, illustrated in Figure \ref{fig:reconfiguration}, reveal that with longer reconfiguration delays, NegotiaToR can still achieve good performance, showing the good generality of its design.

\subsection{Main results}
\label{sec:evaluation_results}

Now that we have investigated NegotiaToR's design elements, we evaluate its overall performance.

\vspace{6pt}

\noindent{\textbf{FCT and goodput on both topologies.}} We compare NegotiaToR's mice flow FCT and goodput with the traffic-oblivious scheme on both topologies. 
Considering real-world deployment practicality, we also show the results when the priority queue for mice flow prioritization is disabled. 

\begin{figure}[t]
    \centering
  
    \begin{minipage}[h]{0.4\textwidth}
        \centering
        \includegraphics[width=\linewidth]{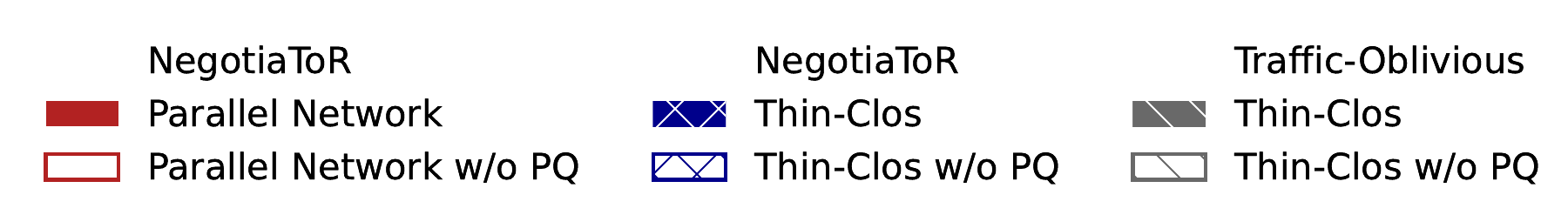}
        \vspace{-0.25in}
    \end{minipage}
  
    \hfill
    \subfigure[Mice flow FCT]{
        \begin{minipage}[b]{0.2\textwidth}
          \centering
          \includegraphics[width=\linewidth]{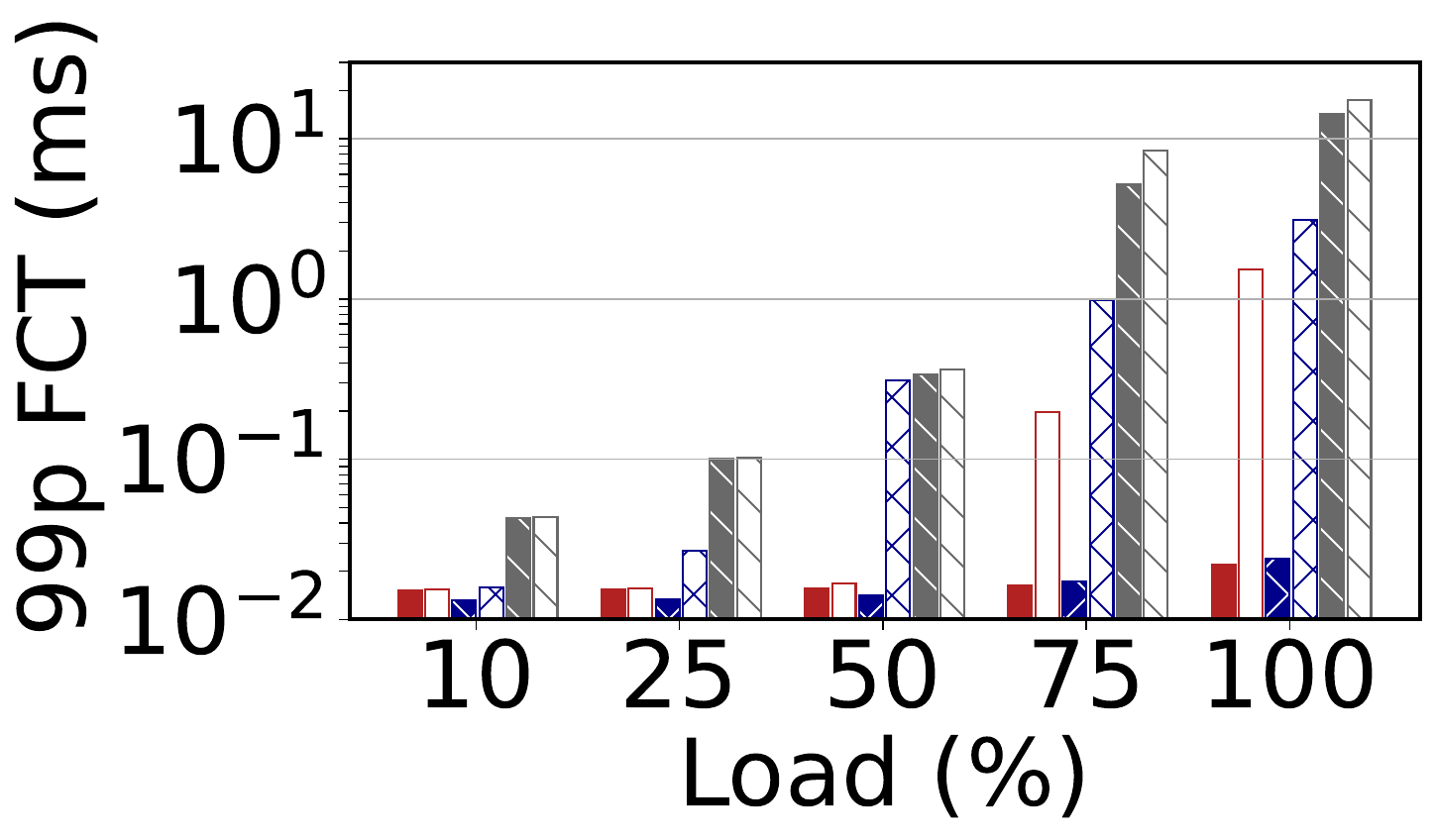}
        \label{fig:base-fct}
        \vspace{-0.2in}
        \end{minipage}
    }
    \hfill
    \subfigure[Goodput]{
        \begin{minipage}[b]{0.2\textwidth}
          \centering
          \includegraphics[width=\linewidth]{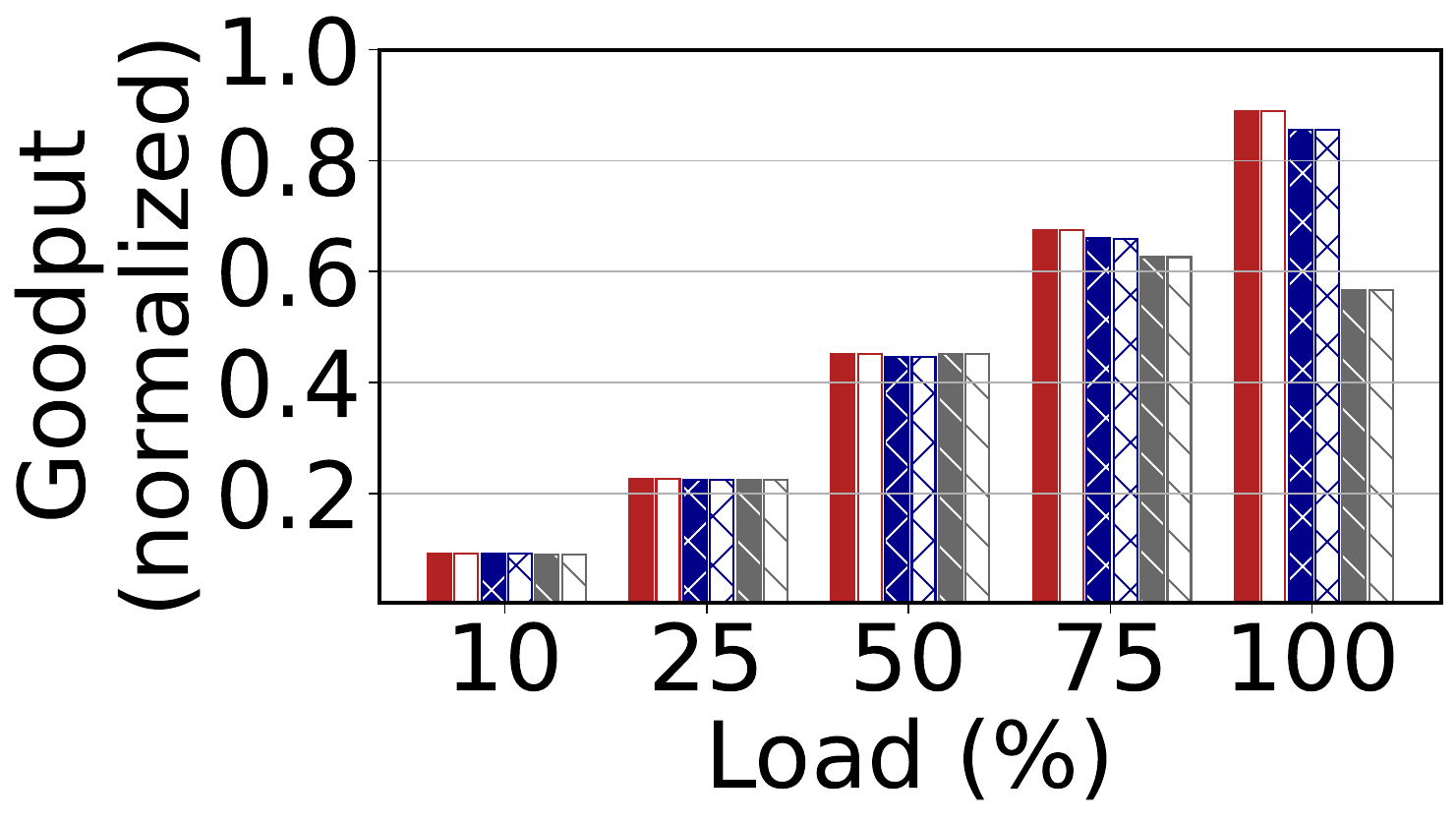}
        \label{fig:base-goodput}
        \vspace{-0.2in}
        \end{minipage}
    }
    \hfill
    \vspace{-5pt}
    \caption{FCT and goodput at various loads. Results with priority queues (PQ) for mice flow prioritization enabled and disabled are both shown.}
    \label{fig:base}
    \vspace{-5pt}
\end{figure}

NegotiaToR achieves better mice flow FCT at all loads regardless of topologies, as in Figure \ref{fig:base-fct}. With priority queues enabled, NegotiaToR's FCT is consistently one to two orders of magnitude better. 
Even without priority queues, such significant advantages still exist at lighter loads. 
This is because, for the traffic-oblivious scheme, the detouring of relayed traffic damages mice flow FCT, especially when elephant flows are spread across the network and block mice flows at intermediate nodes. While for NegotiaToR, the on-demand scheduling together with the scheduling delay bypassing enables prompt transmission of mice flows.

Regarding goodput, NegotiaToR remarkably outperforms the traffic-oblivious scheme on both topologies at high loads, as in Figure \ref{fig:base-goodput}. 
Under low loads, the traffic-oblivious scheme achieves high goodput by utilizing empty links to relay traffic. As the load increases, relayed traffic saturates the network, competing for bandwidth and becoming a bottleneck for goodput. 
In contrast, NegotiaToR's on-demand scheduling can better utilize the network's bandwidth and reduce bandwidth waste.

Through Figure \ref{fig:base}, we also see that no matter which topology is used,  NegotiaToR maintains comparable performance under identical parameter settings, where the performance on the thin-clos topology is marginally lower than on the parallel network due to its limited connectivity. These results underscore that NegotiaToR can adapt well to various flat topologies.

\vspace{6pt}

\begin{figure}
    \centering
    \includegraphics[width=0.43\linewidth]{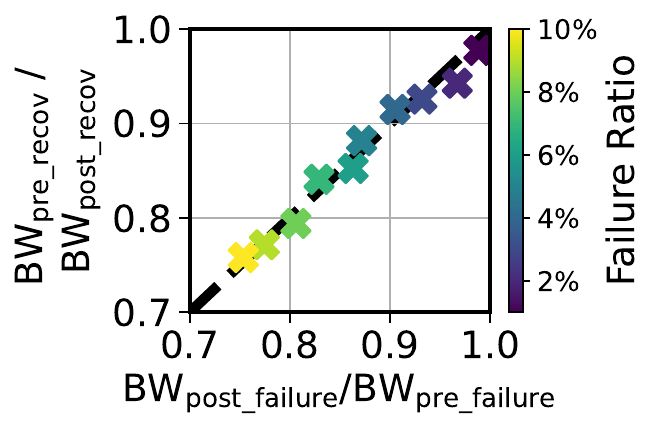}\\
    \vspace{-5pt}
    \caption{NegotiaToR's bandwidth usage changes during link failure and recovery.}
    \label{fig:recovery}
    \vspace{-5pt}
\end{figure}

\noindent{\textbf{Fault tolerance.}} We test the effectiveness of NegotiaToR's fault tolerance mechanism. In the $30ms$ real-world duration, we simulate different levels of simultaneous link failures on the parallel network topology, recover them, and show the bandwidth usage changes in Figure \ref{fig:recovery}. Since a single egress or ingress link failure will affect all traffic passing through it, this will lead to disproportional bandwidth reduction. 
With 1\% of links failing, NegotiaToR's bandwidth usage drops to 98.9\%, while a 10\% failure rate leads to 75.3\%. Upon link recovery, the bandwidth usage returns to its pre-failure level. This validates NegotiaToR's robust fault tolerance capability. 
We also make micro-observations to better understand its behavior under link failures, with results shown in Appendix \ref{sec-appendix2}.

\subsection{Deep dive results}
\label{sec:evaluation_deep_dive}

We further evaluate NegotiaToR under various scenarios to better understand its performance.

\vspace{6pt}

\begin{figure}[t]
    \centering
    \begin{minipage}[h]{0.4\textwidth}
        \centering
        \includegraphics[width=\linewidth]{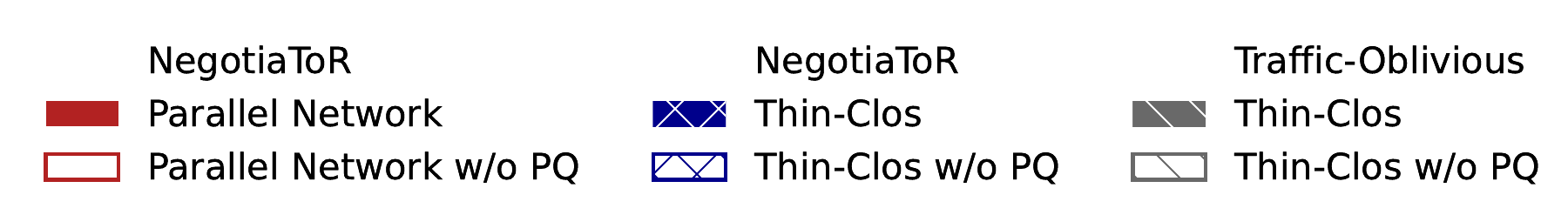}
        \vspace{-0.25in}
    \end{minipage}
  
    \hfill
    \subfigure[Mice flow FCT]{
        \begin{minipage}[b]{0.2\textwidth}
          \centering
          \includegraphics[width=\linewidth]{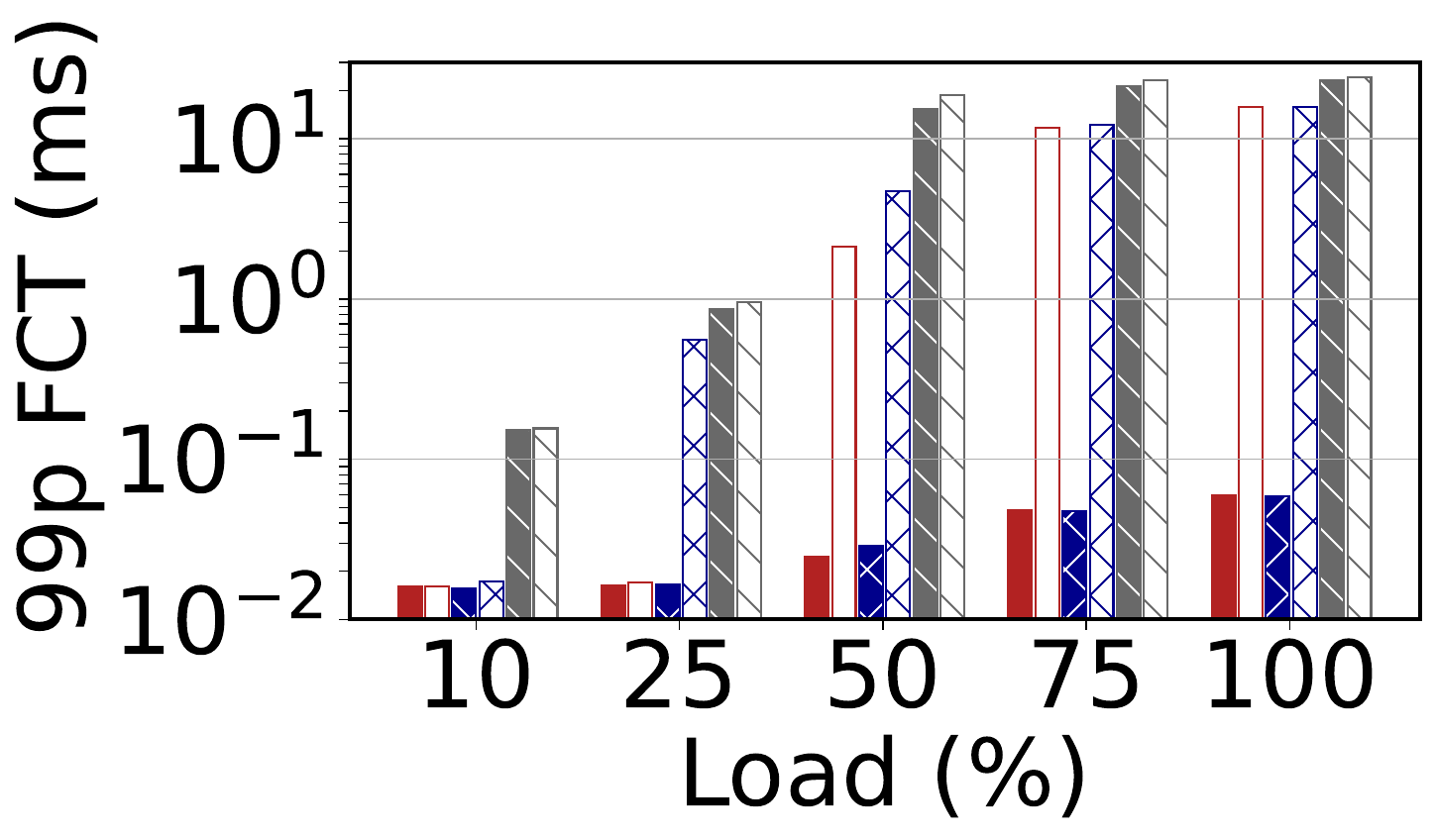}
        \label{fig:no-speedup-fct}
        \vspace{-0.2in}
        \end{minipage}
     }
    \hfill
    \subfigure[Goodput]{
        \begin{minipage}[b]{0.2\textwidth}
          \centering
          \includegraphics[width=\linewidth]{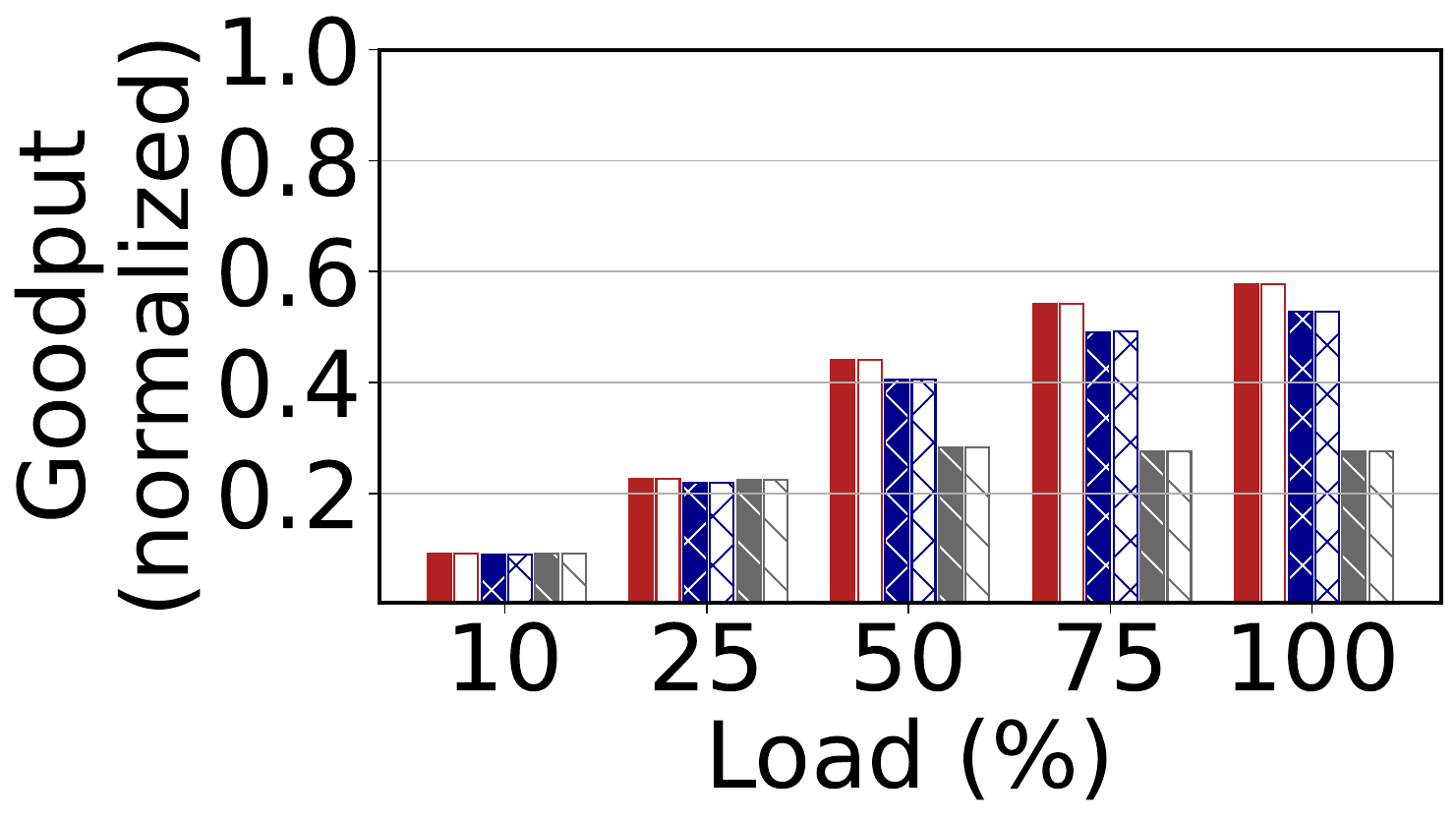}
        \label{fig:no-speedup-goodput}
        \vspace{-0.2in}
        \end{minipage}
    }
    \hfill
    \vspace{-5pt}
    \caption{FCT and goodput at various loads with no speedup.}
    \label{fig:no-speedup}
    \vspace{-5pt}
\end{figure}

\noindent{\textbf{Performance under constrained bandwidth.}}
Previous evaluations are conducted with a $2\times$ speedup. Here we remove this speedup, provide identical bandwidth to ToR uplinks and downlinks, and test NegotiaToR and the traffic-oblivious scheme's performance under the same workload with \S\ref{sec:evaluation_results}.  
Simulations are done on both the parallel network and thin-clos topologies. 
Results illustrated in Figure \ref{fig:no-speedup} align with previous findings. NegotiaToR's on-demand scheduling can better exploit the constrained bandwidth and achieve good performance, highlighting its practicality.

\vspace{6pt}

\begin{figure}[t]
    \centering
  
    \subfigure[Sensitivity of predefined phase timeslot duration: mice flow FCT]{
        \begin{minipage}[b]{0.14\textwidth}
          \centering
          \includegraphics[width=\linewidth]{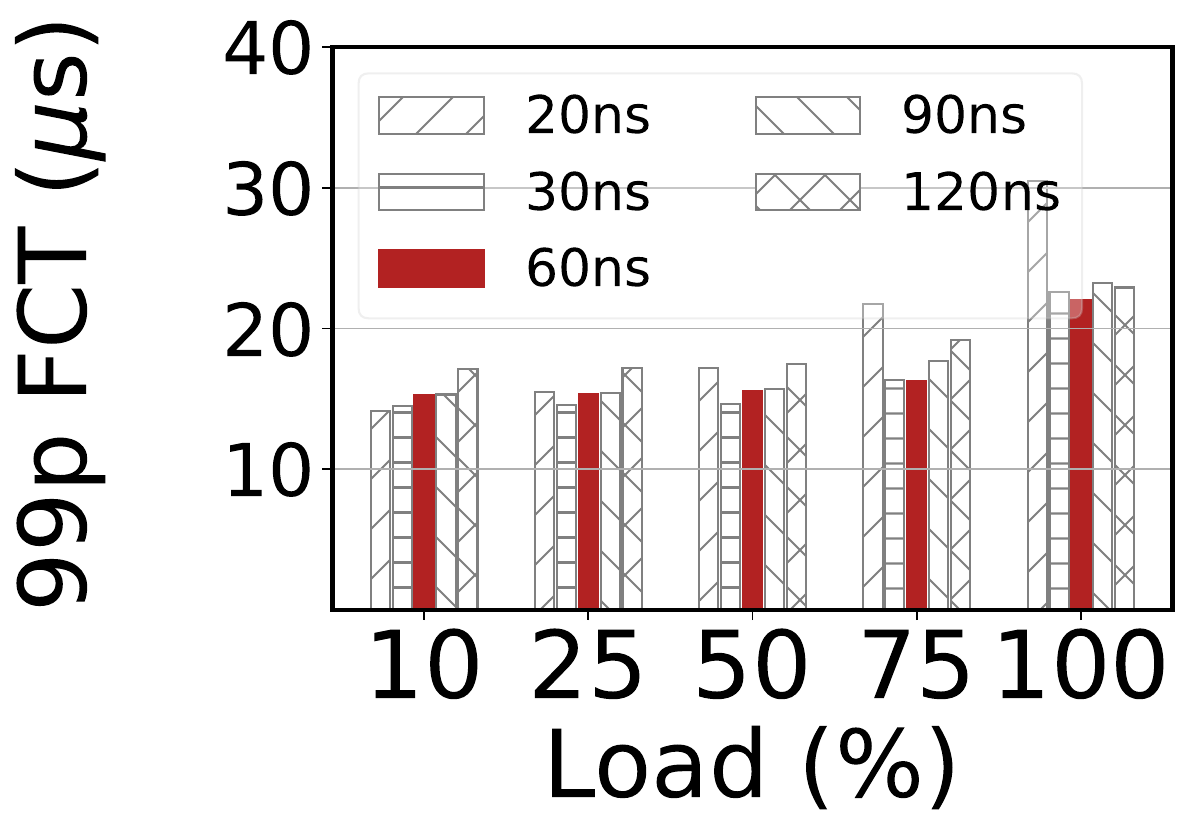}

        \end{minipage}
        \label{fig:parameter-setting-fct-predefined}
        
    }
    \hfill
    \subfigure[Sensitivity of scheduled phase length (in timeslots): mice flow FCT and goodput]{
        \begin{minipage}[b]{0.3\textwidth}
          \centering
          
          \includegraphics[width=0.4667\linewidth]{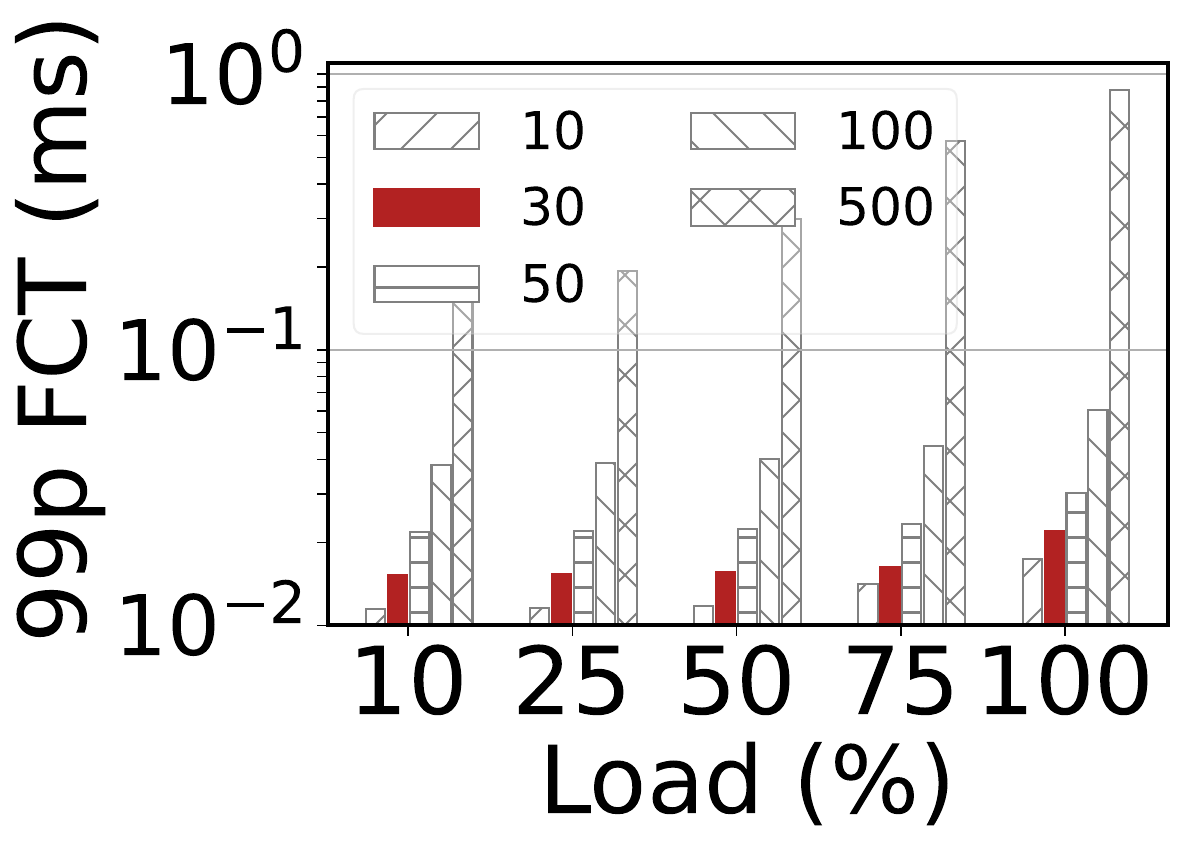}
          \hfill
          \includegraphics[width=0.4667\linewidth]{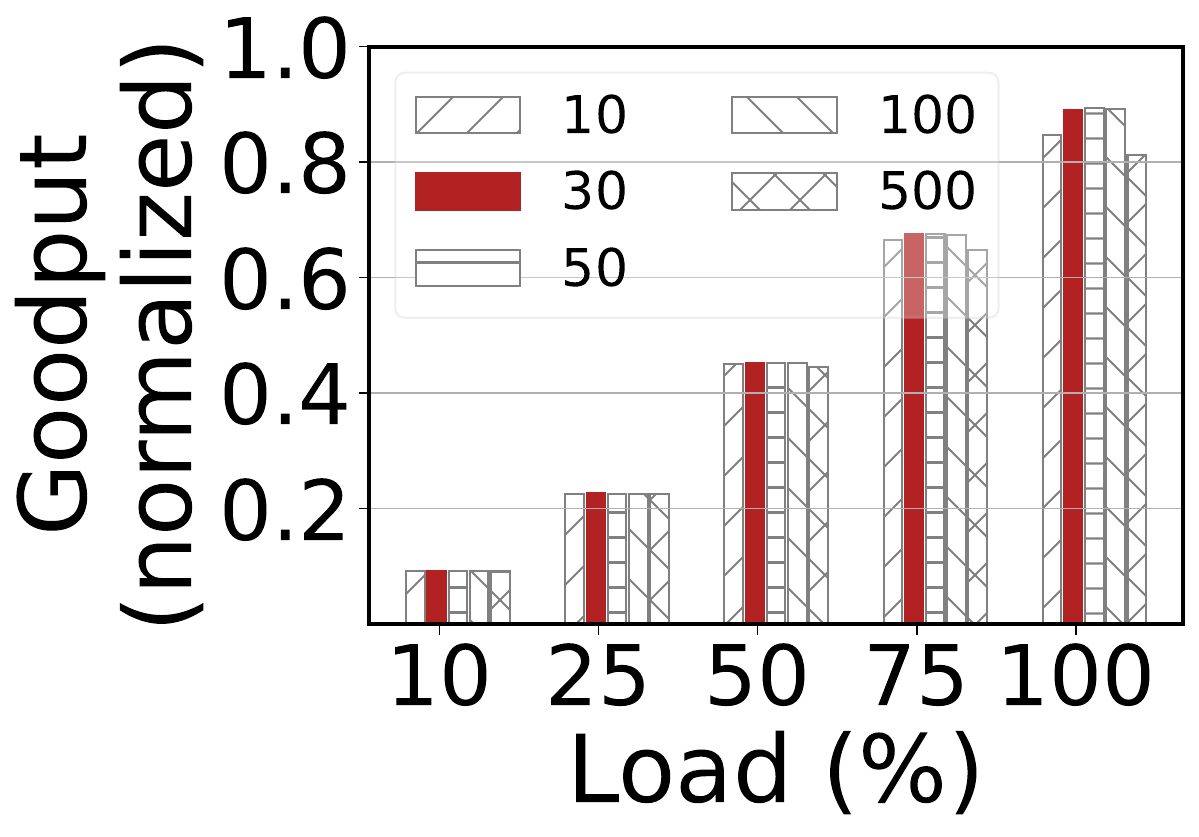}

        \end{minipage}
        \label{fig:parameter-setting-scheduled}
        
    }
    \hfill

    \vspace{-5pt}
    \caption{FCT and goodput under various parameter settings on the parallel network topology. Bars marked in red indicate the setting the evaluation uses by default.}
    \label{fig:parameter-setting}
\end{figure}

\noindent{\textbf{Parameter sensitivity experiment.}}
We investigate the impact of the length of two phases in NegotiaToR by adjusting one at a time. We show experiments conducted on the parallel network under the workload used in \S\ref{sec:evaluation_results} as an example.
We first adjust the duration of each timeslot in the predefined phase, which affects the amount of data that can be piggybacked without scheduling, ranging from 20ns to 120ns including a 10ns guardband. We show the mice flow FCT in Figure \ref{fig:parameter-setting-fct-predefined}, and the goodput is omitted because the difference with Figure \ref{fig:base-goodput} is minor. 
Subsequently, we adjust the length of the scheduled phase from 10 to 500 timeslots, and present the performance in Figure \ref{fig:parameter-setting-scheduled}.
The results match our analysis in \S\ref{sec-epoch-length} well.
They also demonstrate that even when the parameters are approximately set around their optimal values, the impact on NegotiaToR’s performance is minor, showing its robustness.

\vspace{6pt}

\begin{figure}[t]
    \centering
  
    \begin{minipage}[h]{0.43\textwidth}
        \centering
        \includegraphics[width=\linewidth]{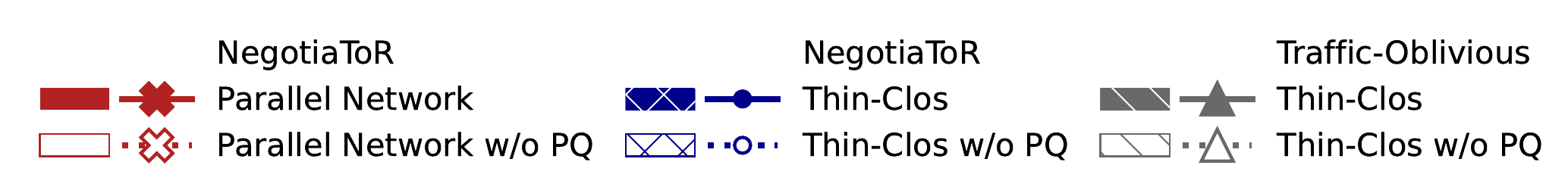}
        \vspace{-0.23in}
    \end{minipage}
    \subfigure[Performance under Hadoop \cite{DBLP:conf/sigcomm/RoyZBPS15} mixed with incast. From left to right is mice flow FCT of the background traffic, average incast finish time, and overall goodput]{
        \begin{minipage}[b]{0.45\textwidth}
        \vspace{-0.05in}
          \centering
          \hfill
          \includegraphics[width=0.29\linewidth]{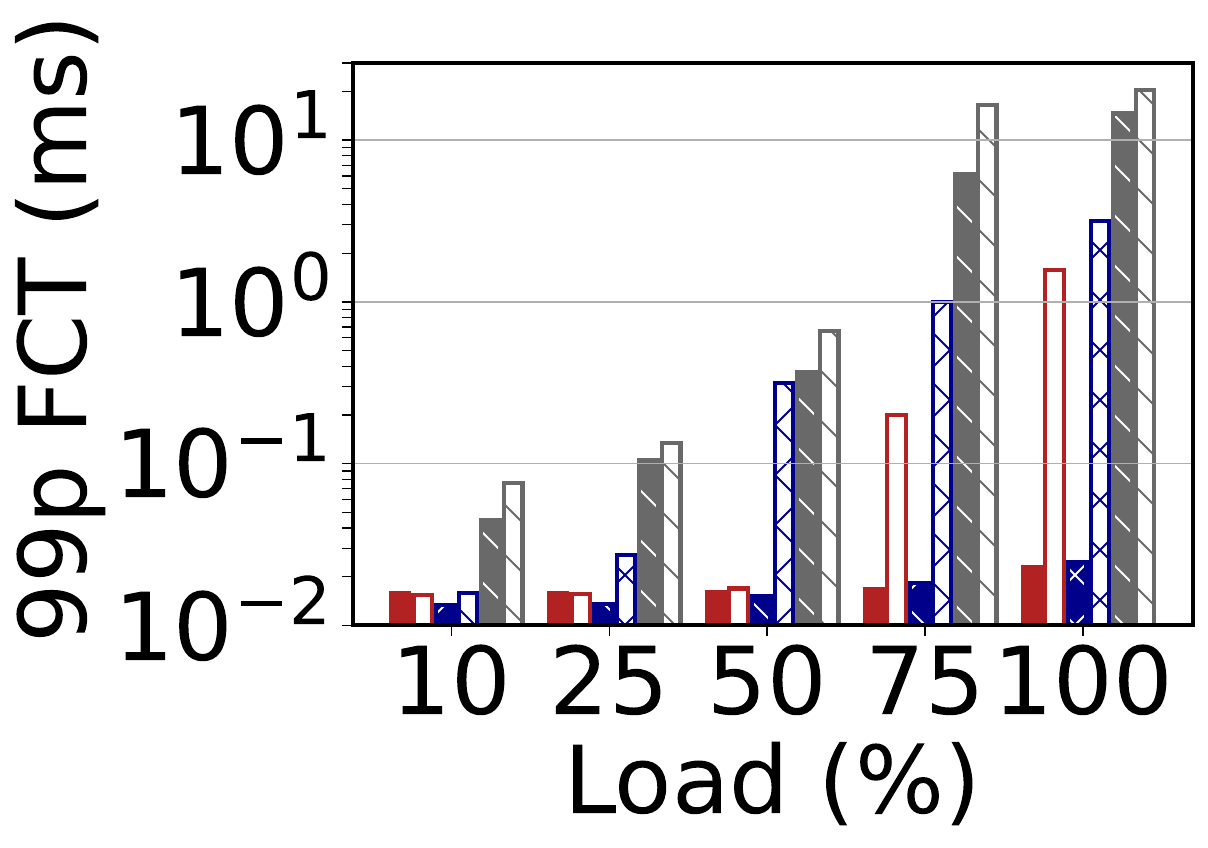}
          \hfill
          \includegraphics[width=0.29\linewidth]{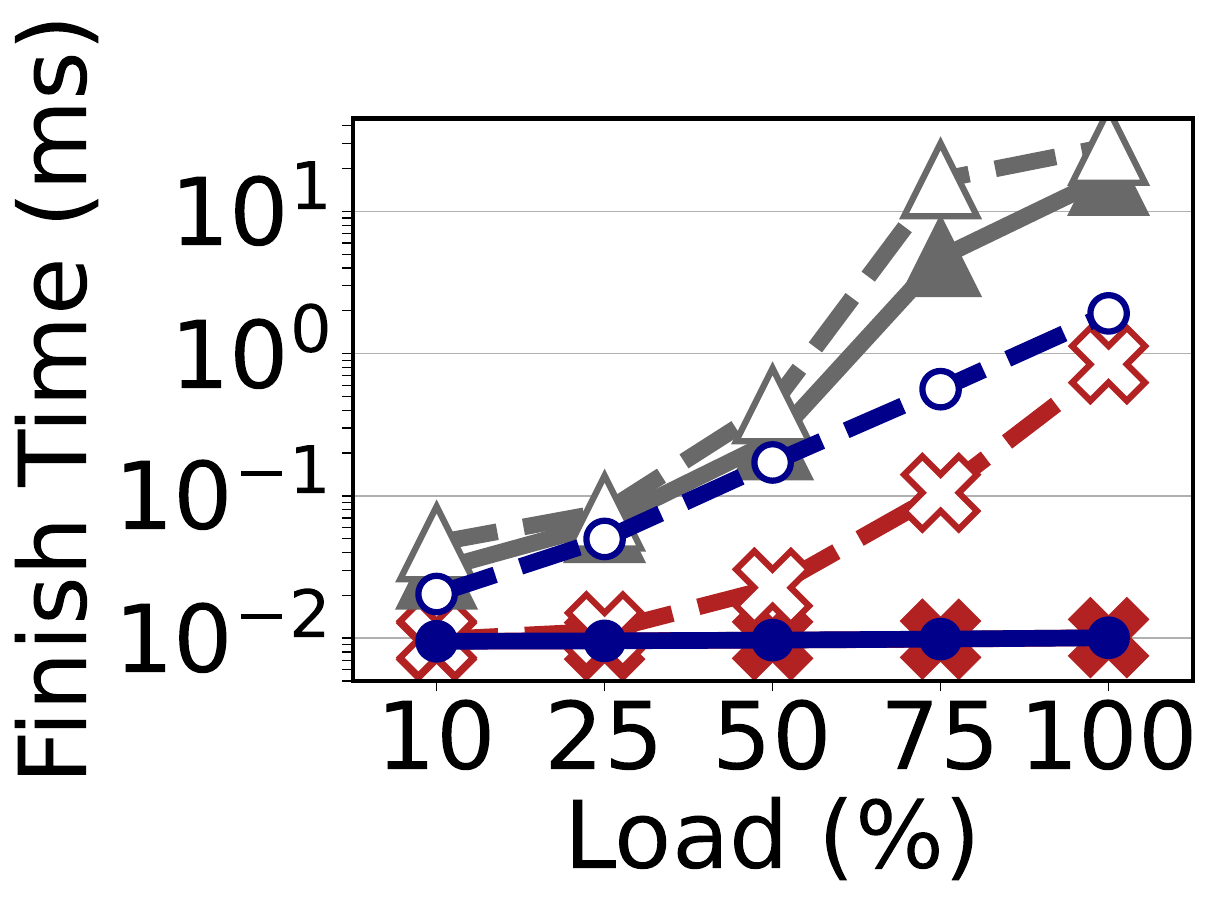}
          \hfill
          \includegraphics[width=0.29\linewidth]{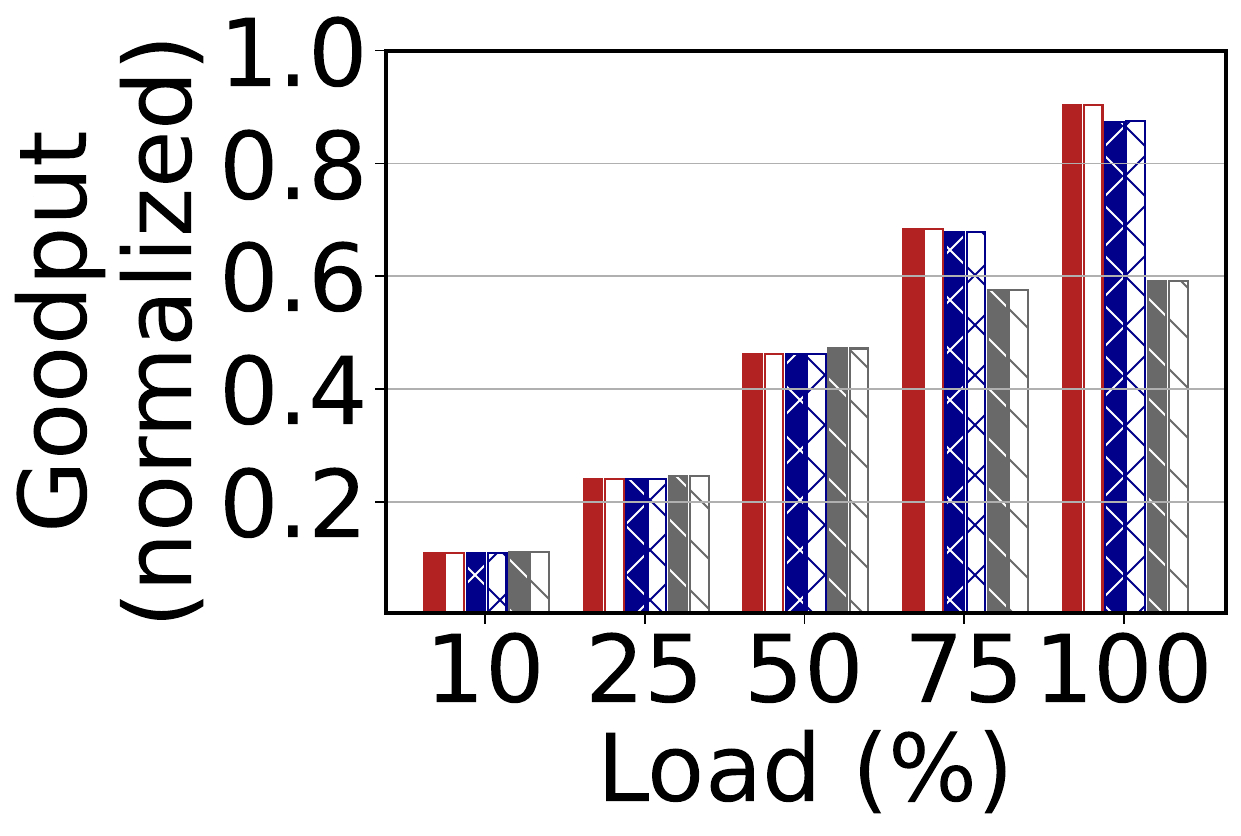}
          \hfill
        \label{fig:mixed-performance}
        \vspace{-0.05in}
        \end{minipage}
    }
    \subfigure[Mice flow FCT and goodput under web search workload \cite{dctcp}]{
        \begin{minipage}[b]{0.45\textwidth}
        \vspace{-0.05in}
          \centering
          \vspace{-0.07in}
          \hfill
          \includegraphics[width=0.4\linewidth]{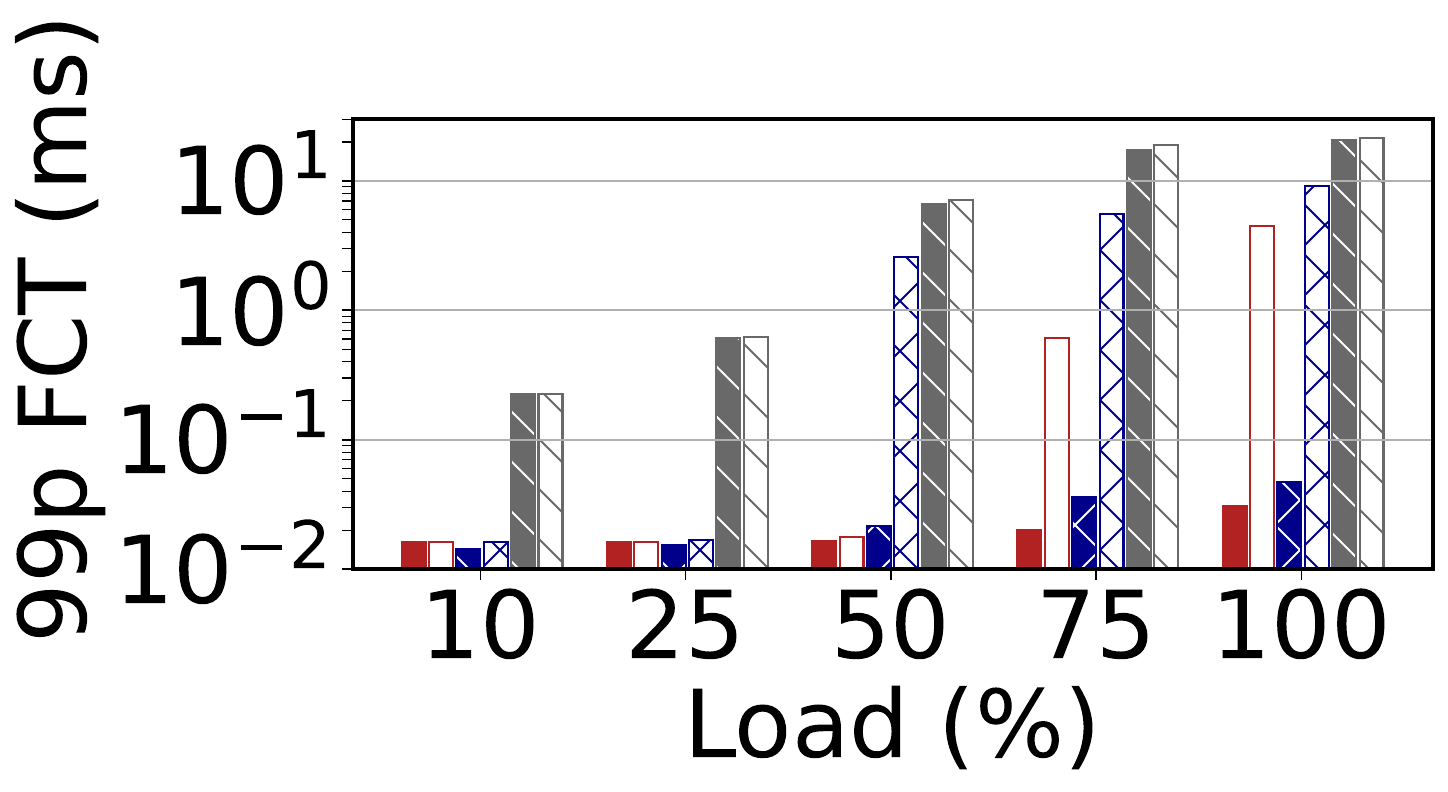}
          \hfill
          \includegraphics[width=0.4\linewidth]{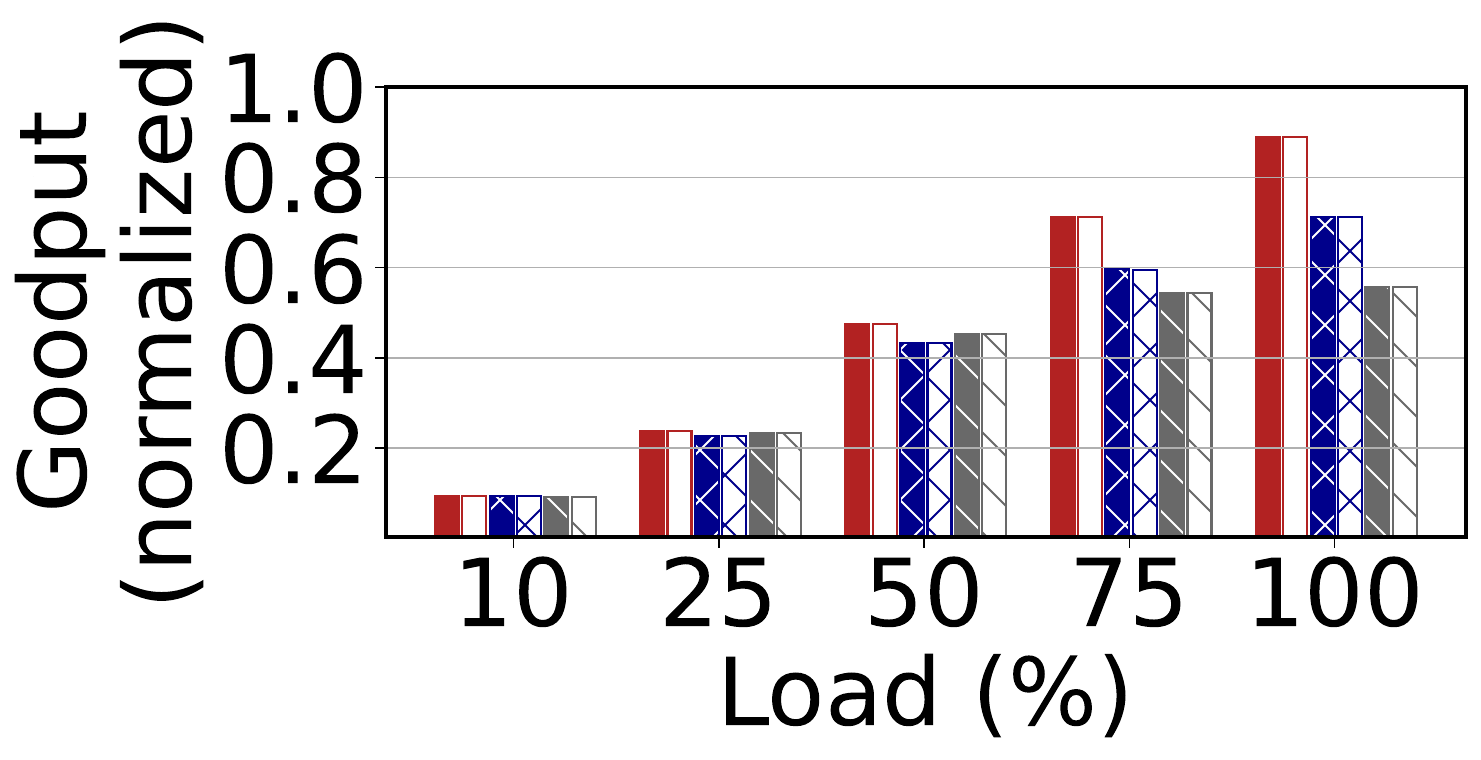}
          \hfill
        \label{fig:web-search-performance}
        \vspace{-0.05in}
        \end{minipage}
    }
    \subfigure[Mice flow FCT and goodput under the aggregated traffic from Google datacenter \cite{GoogleTrace, homa}]{
        \begin{minipage}[b]{0.45\textwidth}
        \vspace{-0.05in}
          \centering
          \vspace{-0.07in}
          \hfill
          \includegraphics[width=0.4\linewidth]{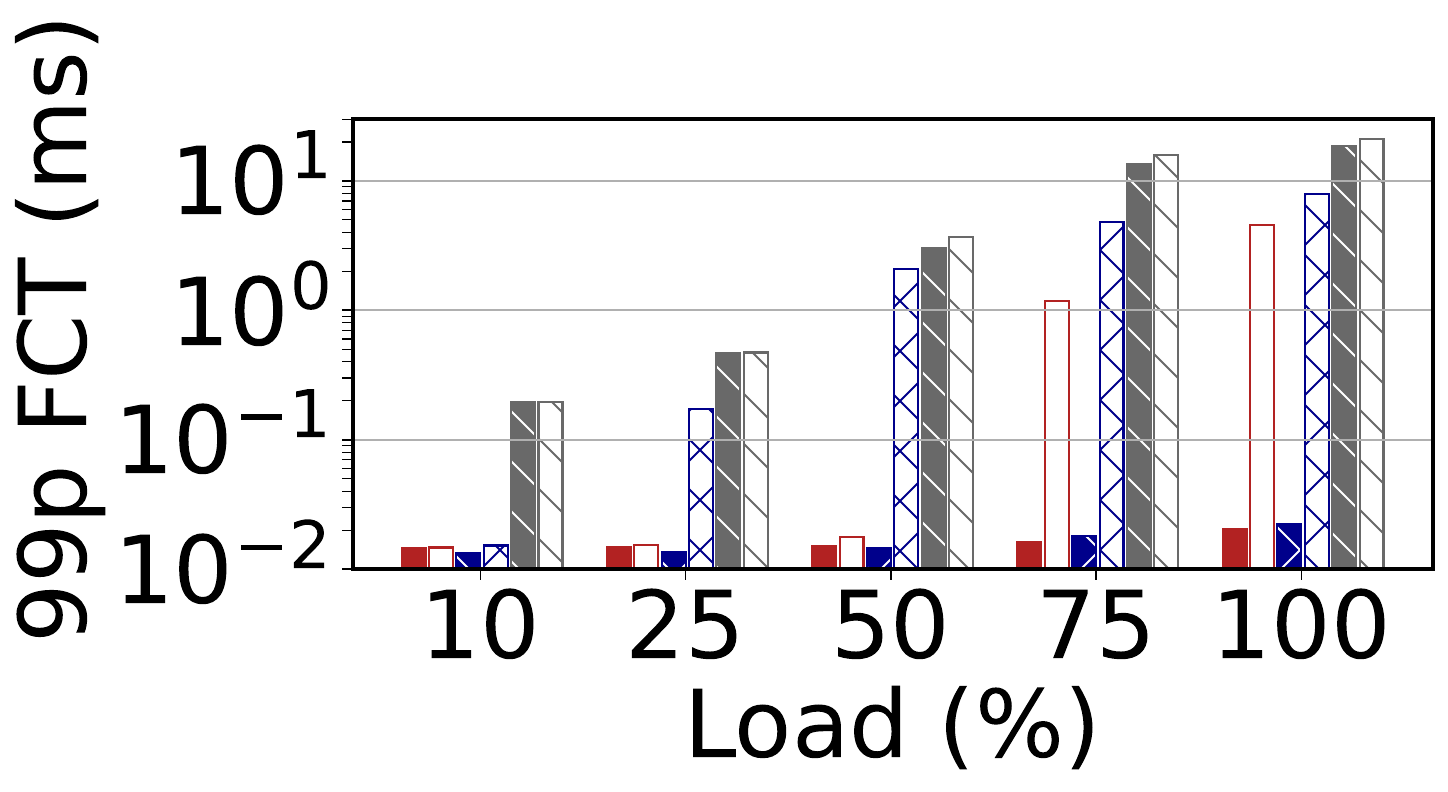}
          \hfill
          \includegraphics[width=0.4\linewidth]{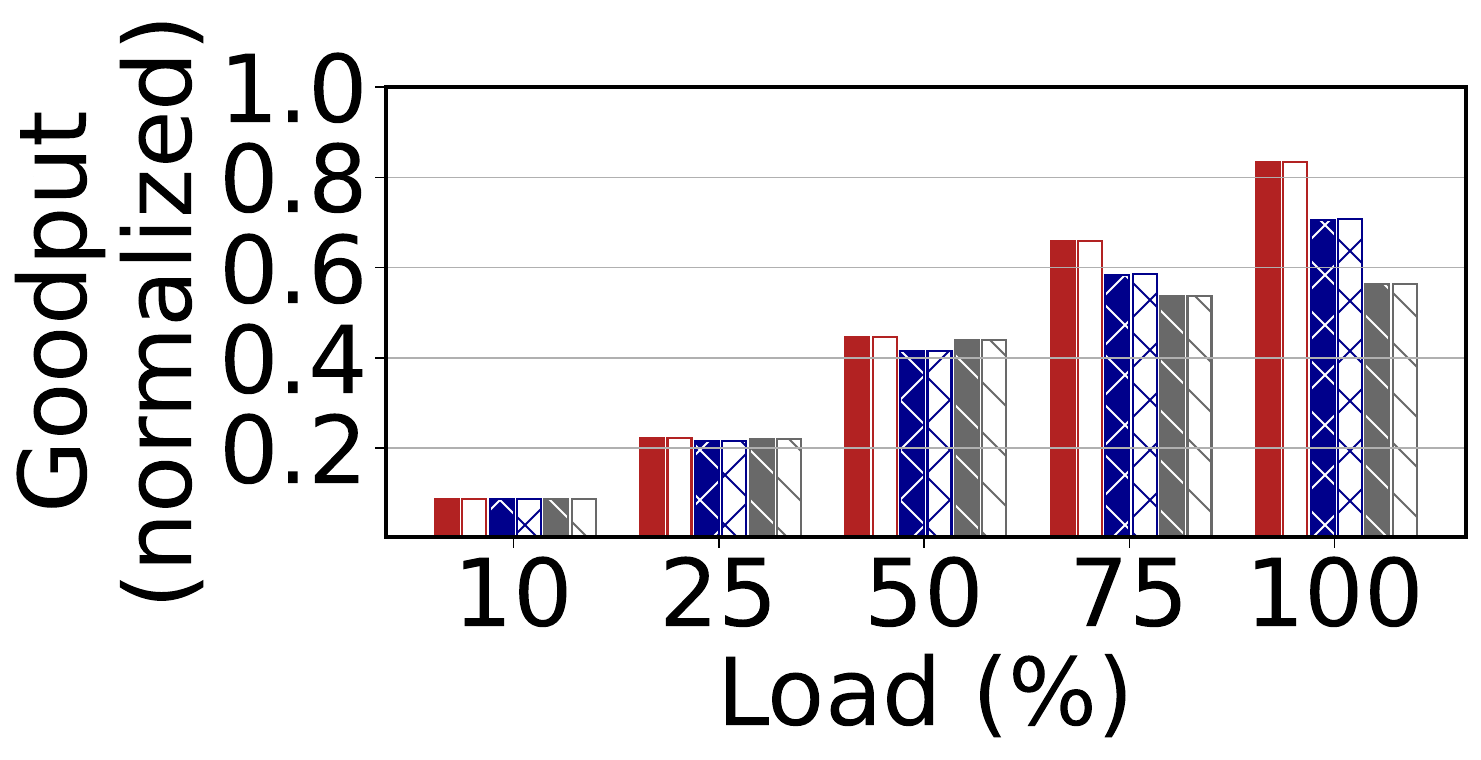}
          \hfill
        \label{fig:google-dcn-performance}
        \vspace{-0.05in}
        \end{minipage}
    }
    \vspace{-7pt}
    \caption{FCT and goodput under more workloads.}
    \label{fig:various-workloads}
    \vspace{-10pt}
\end{figure}

\noindent{\textbf{Performance under various workloads.}}
We further evaluate NegotiaToR under various workloads while maintaining the same epoch length setting as before. 
We first randomly mix incasts on top of the workload used in \S\ref{sec:evaluation_results} to mimic bursty traffic, where each incast has a degree of 20 and a flow size of $1KB$, and all incasts take 2\% of ToR's aggregated downlink bandwidth. Performance of background traffic and incasts are shown in Figure \ref{fig:mixed-performance}. Results indicate that with the incast-optimized scheduling delay bypassing mechanism, NegotiaToR can serve incasts well with minor impact on the background traffic.

Additionally, we test NegotiaToR under two more workloads. The heavier web search \cite{dctcp} workload where more than 80\% flows exceed $10KB$, and the lighter workload from Google datacenter \cite{GoogleTrace, homa} where more than 80\% flows are less than $1KB$. Even though without parameter fine-tuning, results in Figure \ref{fig:web-search-performance} and Figure \ref{fig:google-dcn-performance} show consistent FCT and goodput advantages of NegotiaToR as before, emphasizing the efficiency of its on-demand scheduling.

\section{Related Work}

\noindent{\textbf{Matching algorithms.}} On-demand scheduling on reconfigurable flat topologies can be viewed as a matching problem.
Fast matching algorithms such as PIM \cite{DBLP:journals/tocs/AndersonOST93}, RRM \cite{mckeown1995scheduling} and iSLIP \cite{islip} employ iterations of request, grant, and accept for input-output port connection scheduling in crossbar packet switches.
Unlike the mesh connections of the scheduling logic for ports inside a switch, ToRs in reconfigurable DCNs are multi-ported and interconnected through various topologies, demanding topology-adaptive matching. 
NegotiaToR draws inspiration from them and is uniquely tailored for ToR matching. While maintaining feasibility, it can adapt to various flat topologies. Pipelined scheduling and scheduling delay bypassing are designed to accommodate the long RTT between ToRs.

There are also other practices for using request-grant based or receiver-driven algorithms in DCNs. Packet-switched schemes, like dcPIM \cite{dcpim}, ExpressPass \cite{ExpressPass} and Homa \cite{homa} focus on single-port host-level matching, aiming to control the congestion in the network. They utilize over-commitment to optimize network utilization, contrasting the conflict-free requirement of reconfigurable DCNs.
On-demand scheduling for reconfigurable DCNs has also seen similar approaches. For example, ProjecToR \cite{projector} employs a request-grant based algorithm to schedule optical links among ToRs.
However, it requests at a per-port granularity and leads to minor scalability, and also adds complexity by measuring the waiting delays of bundles of packets at sources for priority decision, whereas NegotiaToR pursues a minimalist approach with binary per-ToR requests and no need for delay measurement. Through experimental explorations including ProjecToR (Appendix \ref{sec:other-design-choices}), we highlight the effectiveness of NegotiaToR's design.

\vspace{6pt}

\noindent{\textbf{Connect ToRs with reconfigurable networks.}}
The networks need to accommodate both short FCT and high goodput with good deployment practicality, serving the dynamic traffic demands between ToRs. 
When the reconfiguration delay of the switching hardware is long (like milliseconds to microseconds) compared with RTT, FCT is a main concern. Proposals use a hybrid packet-switched network \cite{rotornet, helios, c-through, reactor, firefly} or multi-hop routing \cite{firefly, OSA, opera, projector} to optimize mice flow FCT. 
For instance, Opera \cite{opera} reconfigures its topology to a set of pre-calculated expander graphs, enabling mice flows to be immediately sent through multi-hop paths. 
This approach allows Opera to deliver low mice flow FCT in the order of tens of microseconds with the optical network alone. However, due to hardware reconfiguration delay limitations, its direct connections still mismatch the real-time traffic demands, posing challenges in accommodating the dynamic traffic.

With recent advancements in fast switching hardware featuring nanoseconds reconfiguration delay, opportunities have been seen to serve the dynamic traffic.
Previous on-demand scheduling proposals like PULSE \cite{PULSE} ensure performance, but raise scalability concerns due to complex scheduling logic.
Sirius \cite{sirius} utilizes the traffic-oblivious method combined with data relay, which, although simple, encounters difficulties in maintaining high goodput under heavy loads due to bandwidth competition caused by relayed traffic, and in providing short FCT due to the relay latency. 
In contrast, NegotiaToR aims to achieve on-demand scheduling over fast switching hardware through a minimalist design. With distributed matching, pipelined scheduling and incast-optimized scheduling delay bypassing schemes, NegotiaToR offers short FCT and high goodput through one-hop transmission while maintaining low complexity, further exploiting the fast reconfiguration capability.

\vspace{-2pt}
\section{Conclusion}

We presented NegotiaToR, an on-demand reconfigurable DCN architecture with a simple design.
With the two-phase epoch, it runs NegotiaToR Matching distributedly in-band to reconfigure the network according to dynamic real-time traffic demands. 
It also provides an incast-optimized scheduling delay bypassing scheme to mitigate the impact of scheduling delays. 
NegotiaToR is compatible with prevalent flat topologies, and is tailored towards a minimalist design for on-demand reconfigurable DCNs, enhancing practicality. 
By exploiting the fast optical switching technology to provide high performance with low complexity, we hope that NegotiaToR can facilitate the development of next-generation reconfigurable DCNs.

\begin{acks}
We thank our shepherd, Alex Snoeren, and the anonymous SIGCOMM reviewers for their useful feedback on this paper. Yong Cui (cuiyong@tsinghua.edu.cn) is the corresponding author. 
This work was supported by the National Natural Science Foundation of China under Grants 62132009, 62221003 and 62272292.
\end{acks}

\bibliographystyle{ACM-Reference-Format}
\bibliography{reference}

\appendix
\newpage
\section{Appendix}
\label{sec-appendix}

\textit{Appendices are supporting material that has not been peer-reviewed.}

\subsection{Efficiency analysis validation of NegotiaToR Matching}
\label{appendix-negotiator-matching}

\begin{figure}[t]
    \centering
    \subfigure[Parallel network]{
        \begin{minipage}[b]{0.22\textwidth}
            \centering
            \includegraphics[width=\linewidth]{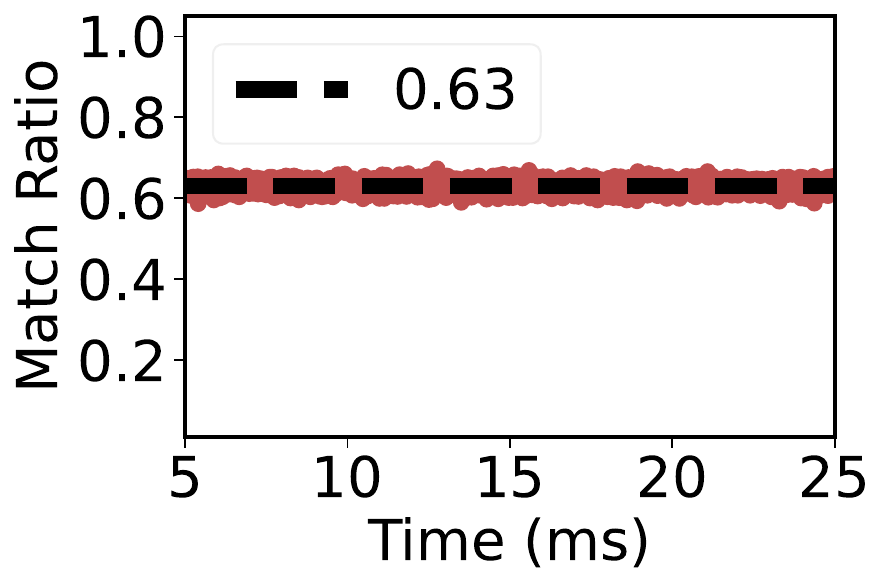}
        \end{minipage}
    }
    \subfigure[Thin-clos]{
        \begin{minipage}[b]{0.22\textwidth}
            \centering
            \includegraphics[width=\linewidth]{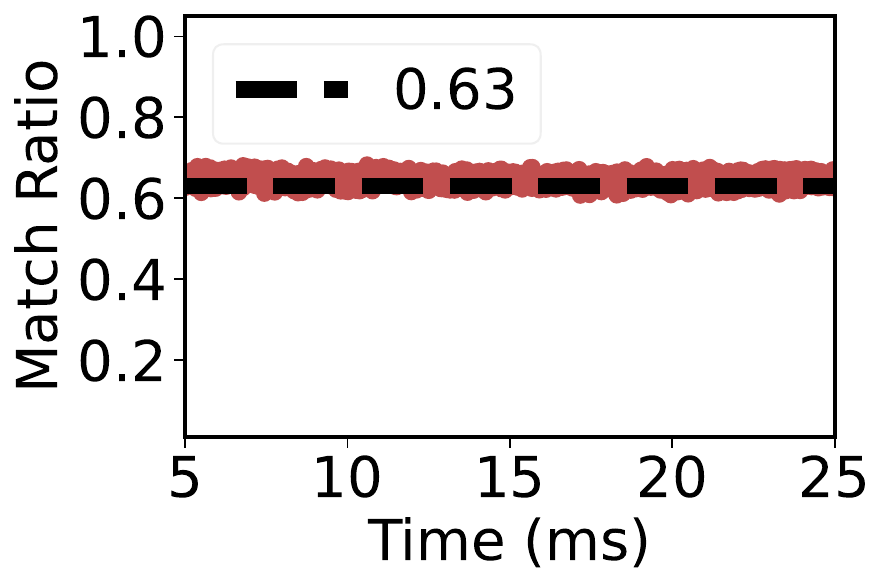}
        \end{minipage}
    }
    \caption{NegotiaToR's match ratio for epochs in both topologies, recorded in the simulation at 100\% load in \S\ref{sec:evaluation_results}.}
    \label{fig:match_ratio}
\end{figure}

To validate the theoretical results of NegotiaToR Matching's efficiency in \S\ref{sub-sec:matching-efficiency}, we examine the scheduling efficiency in our large-scale simulations at 100\% load in \S\ref{sec:evaluation_results}, since heavy-load scenarios are closer to the intense-competition scenario we assumed in the theoretical model. For each epoch, we record the ratio of accepts and grants (we call it match ratio) and depict it in Figure \ref{fig:match_ratio}. Results show that the actual match ratio is consistent with our expectations, with thin-clos topology ($n = 16, E[Y]=0.644$) achieving a slightly higher match ratio than the parallel network topology ($n = 128, E[Y]=0.634$).

\subsection{Details of the experimental exploration of other design choices}
\label{sec-appendix1}

NegotiaToR's design is tailored towards simplicity for practicality. 
To find out whether a little more complexity can bring a significant performance gain, we explored some other design choices. In \S\ref{sec:algorithm_design_discussion}, we found that iterative scheduling, traffic-aware selective relay, informative requests, and stateful scheduling may not provide proportionate performance gains. We give the details of our exploration in this section. For simplicity, we only show the results of the parallel network topology, except for traffic-aware selective relay, which is designed for the thin-clos topology.
Unless specified, simulation follows the default settings in \S\ref{sec:evaluation_setup}.

\subsubsection{Iterative NegotiaToR Matching}
\label{sec:iterative-negotiator-matching}

The impact of introducing iteration to NegotiaToR Matching is twofold.
On one hand, iteration could enhance matching efficiency by fully utilizing unmatched ports, potentially improving goodput. 
However, in optical switching, lower goodput can often be effectively compensated by speedup (i.e., providing more aggregated uplink bandwidth than aggregated downlink bandwidth).
On the other hand, introducing iteration significantly increases the scheduling delay due to the long RTT between ToRs, and also adds complexity. Traffic demand information is more likely to be outdated by the time the scheduling decision is received, since data may have already been sent by previous epochs during the long scheduling process, leading to link waste and thus performance degradation.
Therefore, although the scheduling result may be closer to a maximal match, FCT and goodput could be adversely affected.

For investigation, we design an iterative version of NegotiaToR Matching. After deriving the \textit{accept}, instead of directly sending data, new \textit{request} is sent to destinations again along with indices of unmatched ports for further iterations. 
Multiple rounds of iteration also enlarge the scheduling delay. For one more iteration, the scheduling delay is enlarged by three epochs. 

\begin{figure}
    \centering
    \begin{minipage}[h]{0.4\textwidth}
        \centering
        \includegraphics[width=\linewidth]{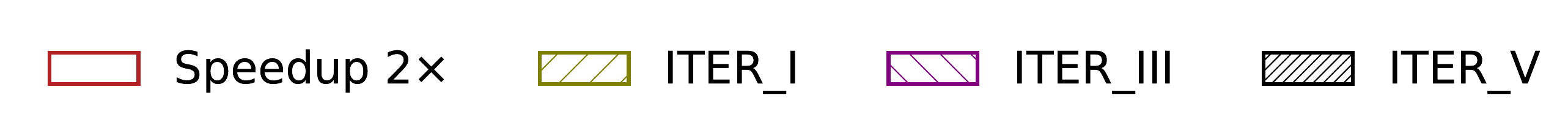}
        \vspace{-0.25in}
    \end{minipage}
    \subfigure[Mice flow FCT]{
        \begin{minipage}[b]{0.22\textwidth}
          \centering
          \includegraphics[width=\linewidth]{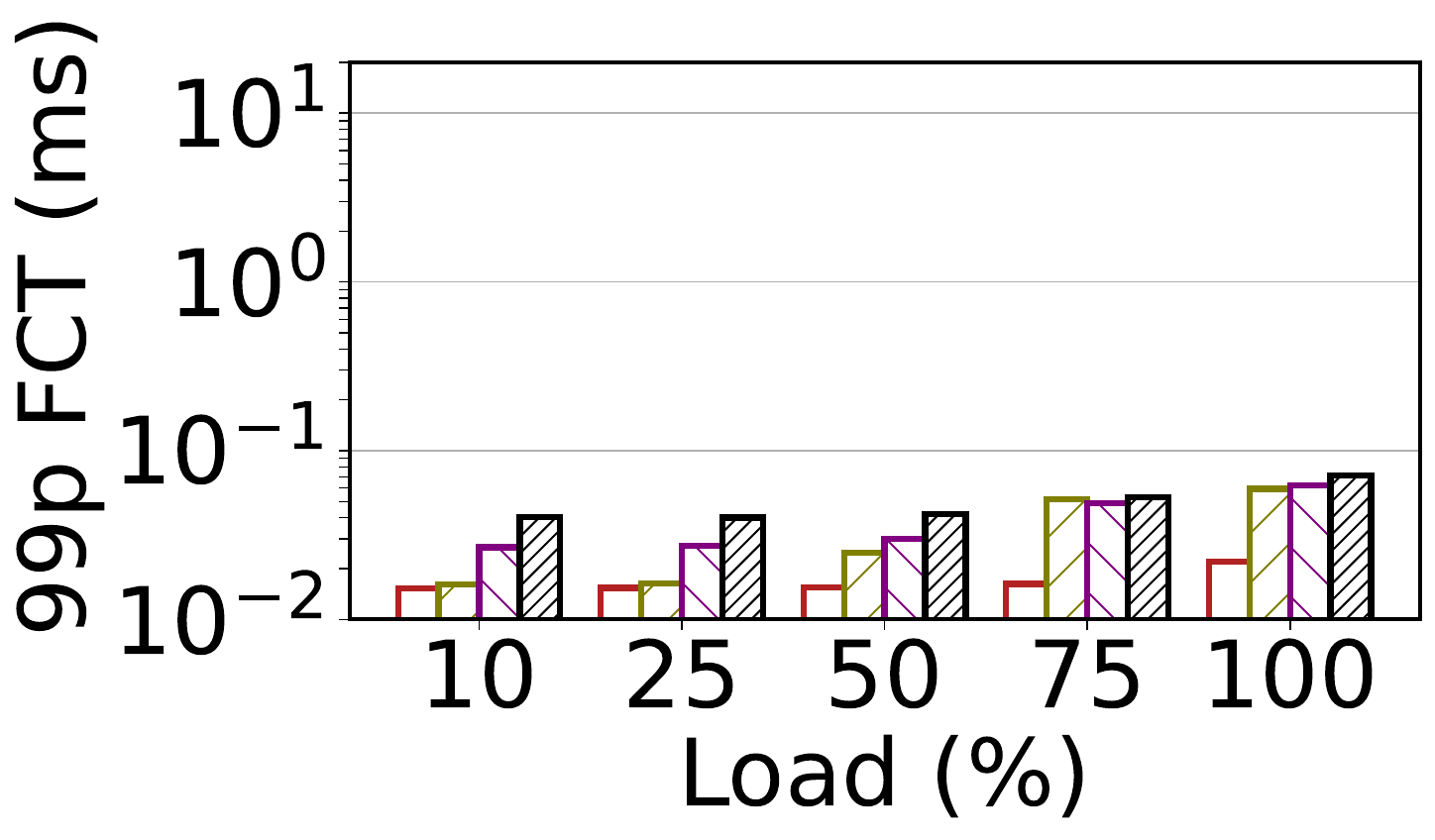}
        \end{minipage}
    }
    \subfigure[Goodput]{
        \begin{minipage}[b]{0.22\textwidth}
          \centering
          \includegraphics[width=\linewidth]{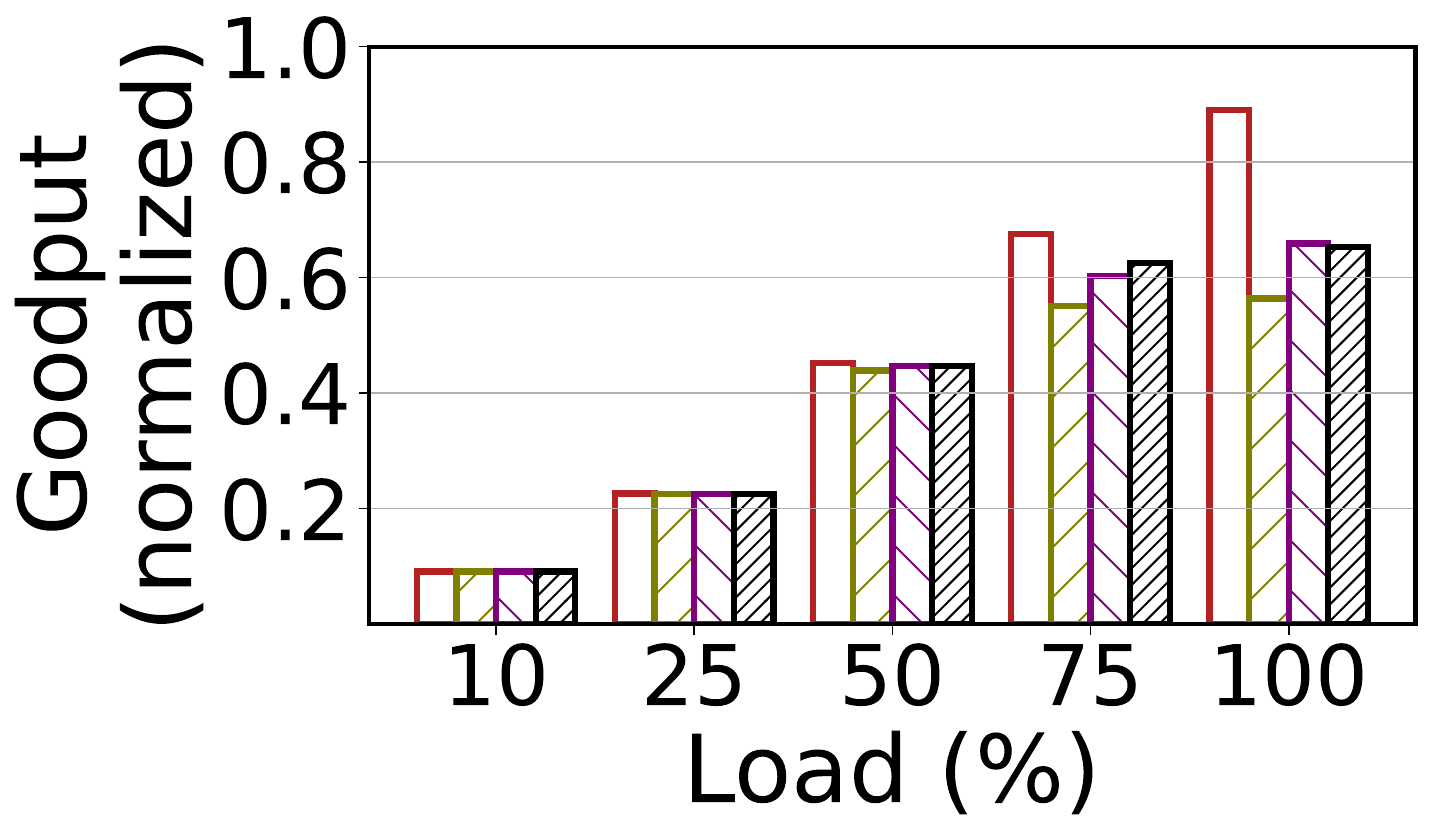}
        \end{minipage}
    }
    \caption{FCT and goodput on the parallel network topology at various loads, with $2\times$ speedup or iteration enabled.}
    \label{fig:iteration}
  \end{figure}

\vspace{6pt}

\noindent{\textbf{Simulation results.}} We conduct simulations of both an iterative and a non-iterative original version of NegotiaToR Matching on the parallel network topology. 
For the non-iterative one, we set the speedup factor to 2, as depicted in \S\ref{sec:evaluation_setup}. For the iterative version, we try 3 and 5 rounds of iterations with no speedup.
We set the size of scheduling messages (indicated by packet header size in simulation) to be the same for fair comparison, even though the iterative algorithm actually needs extra bits to transmit port-level occupation information, which leads to extra overhead.

Results in Figure \ref{fig:iteration} show that, as we expected, iterative scheduling adversely affects FCT at all loads because of the long scheduling delay. Meanwhile, with more rounds of iteration, goodput starts to decrease due to the outdated traffic demand information. 
Notably, at all loads, goodput is consistently equal to or lower than the non-iterative version with 2$\times$ speedup.
This implies that instead of introducing iteration, it is more effective to increase the speedup factor to compensate for the possibly low matching efficiency, which also improves the FCT performance.
Therefore, we conclude that iteration is not an optimal choice for NegotiaToR Matching.

\subsubsection{Traffic-aware selective relay}
\label{sec:traffic-aware-intermediation}

In NegotiaToR, all data is sent through one-hop paths. Avoiding traffic-oblivious relay helps NegotiaToR avoid goodput damages under heavy loads caused by the doubling of data volume, as well as mice flow FCT damages caused by more hops. Meanwhile, data relay requires congestion control mechanisms to avoid buffer overflow at intermediate nodes, adding complexity, which NegotiaToR does not need. 

A third design choice that is potentially beneficial for goodput also exists, where data relay is enabled only for elephant flows under light loads \cite{trod}. 
In the parallel network topology (Figure \ref{fig:negotiator_topology_big_switch}), each ToR pair can transmit data through all ports, making relaying unnecessary. 
However, consider the thin-clos topology (Figure \ref{fig:negotiator_topology_thin_clos}), where each pair of ToRs is connected by a single port-to-port path. By enabling data relay, the number of available paths between ToRs is increased, and data transmission between ToRs could be done through multiple ports simultaneously, which could potentially improve goodput.

To observe the impact of such data relay on NegotiaToR, we extend NegotiaToR to support traffic-aware selective relay, aiming to improve the link utilization of the thin-clos topology when empty links are available, as we discussed in \S\ref{sec:algorithm_design_discussion}. For elephant flows, in addition to direct data transmission, it also considers utilizing lightly loaded ToRs to relay the data. The limit on flow size is to protect mice flows from increased FCT, while the constraints of the intermediate ToR are to avoid goodput reduction caused by competition with directly transmitted data.

\begin{figure}[t]
    \includegraphics[scale=0.43]{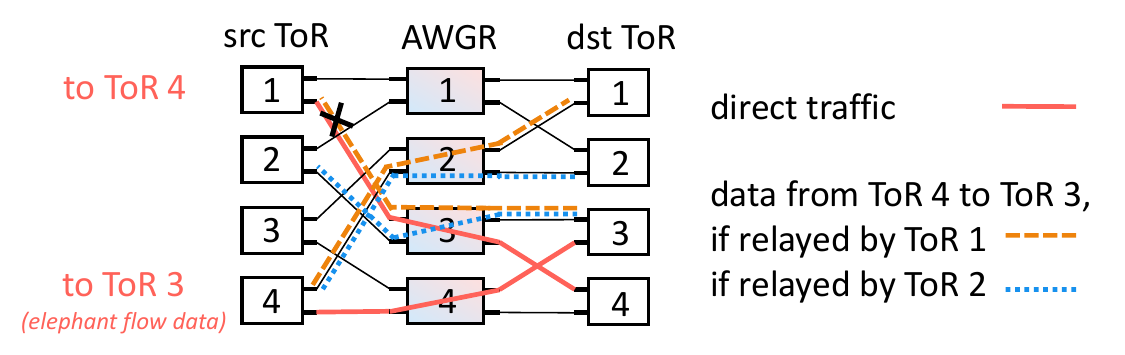}
    \caption{Traffic-aware selective relay on the thin-clos topology. Data relay is enabled for the large volume of data from ToR 4 to ToR 3. When selecting the intermediate ToR, ToR 1 is excluded due to the high volume of direct traffic to ToR 4, which would result in bandwidth competition at ToR 1 $\rightarrow$ AWGR 3 link. In contrast, ToR 2 can be selected to relay the data, avoiding potential goodput damage.}
    \label{fig:thin-clos-intermediation-example}
\end{figure}

We integrate traffic-aware selective relay into NegotiaToR Matching, including selecting the data to be relayed, filtering out loaded links using local direct traffic information, and preventing buffer overflow at the intermediate ToR with congestion control. We give an illustrative example in Figure \ref{fig:thin-clos-intermediation-example}.

\noindent\textcircled{1} Before sending requests, source ToRs examine the data volume in the per-destination queue, and determine potential intermediate candidates.

\begin{itemize}
    \item Co-designed with the mice flow prioritization mechanism (\S\ref{sec:mice_flow_prioritization}), we disable relay for data in the highest-priority queue (i.e., mice flow data), and only enable it for data in the lowest-priority queue (i.e., elephant flow data) if the data volume exceeds a certain threshold. 
    This ensures that mice flows are always sent directly without FCT concerns, while elephant flows are relayed only when they have enough data to fill extra links so that they can finish sooner than if they were sent directly. 
    \item Source ToRs then use local traffic information and find intermediate ToR candidates, excluding any that would cause bandwidth competition due to shared links with high-volume direct traffic. Requests for data relay are sent to the available candidates.
\end{itemize}

\noindent\textcircled{2} In the GRANT step, the candidates evaluate their intermediate queue lengths and high-volume direct traffic conflicts on shared links. 
If the queue still has spare capacity, and there exists no high-volume concurrent direct traffic, then the candidates grant the request, indicating the maximal data volume it can relay.

\noindent\textcircled{3} Finally, in the ACCEPT step, the transmission of direct traffic is prioritized over relayed traffic. After sending direct data first, the source ToR then sends the data to be relayed from the lowest-priority queue to the intermediate ToR.

\begin{table}[t]
    \resizebox{\columnwidth}{!}{
    \begin{tabular}{|l|ccccc|}
    \hline
    \multirow{2}{*}{} & \multicolumn{5}{c|}{99p Mice Flow FCT ($\mu$s) / Normalized Goodput}                                                                                   \\ \cline{2-6} 
                        & \multicolumn{1}{c|}{10\%} & \multicolumn{1}{c|}{25\%} & \multicolumn{1}{c|}{50\%} & \multicolumn{1}{c|}{75\%} & 100\% \\ \hline
    Base              & \multicolumn{1}{c|}{13.2/9.1\%}     & \multicolumn{1}{c|}{13.4/22.5\%}     & \multicolumn{1}{c|}{14.2/44.6\%}     & \multicolumn{1}{c|}{17.3/66.0\%}     &   23.8/85.6\%    \\ \hline
    Two-Hop    & \multicolumn{1}{c|}{13.4/9.1\%}     & \multicolumn{1}{c|}{14.0/22.6\%}     & \multicolumn{1}{c|}{16.8/45.1\%}     & \multicolumn{1}{c|}{19.2/66.9\%}     &   24.2/86.8\%    \\ \hline
    \end{tabular}%
    }
    \vspace{0.1in}
    \caption{FCT and goodput on the thin-clos topology at various loads, when the traffic-aware selective relay is enabled.}
    \label{table:intermediation}
\end{table}

\vspace{6pt}

\noindent{\textbf{Simulation results.}}
We conduct simulations of NegotiaToR with traffic-aware selective relay enabled on the thin-clos topology. We show the results in Table \ref{table:intermediation} under the optimal relay setting we found. When data relay is enabled, FCT is barely affected due to we only selectively intermediating elephant flows.
However, goodput is also merely improved. This is because, at light loads, NegotiaToR, with a $2\times$ speedup, already achieves \textit{near-optimal} goodput considering the overhead introduced by guardbands and headers, leaving little room for improvements. While at heavy loads where links are already saturated, data relay does not help, and is turned off by our traffic-aware relay scheme for most of the time. Therefore, we do not employ data relay in NegotiaToR.

\subsubsection{Informative requests}
\label{sec:informative-requests}

    \begin{table}[t]
        \centering
        \resizebox{\columnwidth}{!}{
        \begin{tabular}{|l|ccccc|}
        \hline
        \multirow{2}{*}{} & \multicolumn{5}{c|}{99p Mice Flow FCT ($\mu$s) / Normalized Goodput}                                                                                   \\ \cline{2-6} 
                          & \multicolumn{1}{c|}{10\%} & \multicolumn{1}{c|}{25\%} & \multicolumn{1}{c|}{50\%} & \multicolumn{1}{c|}{75\%} & 100\% \\ \hline
        Base              & \multicolumn{1}{c|}{15.3/9.1\%}     & \multicolumn{1}{c|}{15.4/22.6\%}     & \multicolumn{1}{c|}{15.6/45.2\%}     & \multicolumn{1}{c|}{16.3/67.5\%}     & 22.0/89.0\%  \\ \hline
        Data-Size         & \multicolumn{1}{c|}{15.6/9.1\%}     & \multicolumn{1}{c|}{15.9/22.6\%}     & \multicolumn{1}{c|}{16.4/45.2\%}     & \multicolumn{1}{c|}{23.0/67.6\%}     &   44.2/89.8\%    \\ \hline
        HoL-Delay         & \multicolumn{1}{c|}{15.2/9.1\%}     & \multicolumn{1}{c|}{15.2/22.6\%}     & \multicolumn{1}{c|}{15.3/45.2\%}     & \multicolumn{1}{c|}{15.3/67.6\%}     &   15.5/89.2\%    \\ \hline
        \end{tabular}%
        }
        \vspace{0.1in}
        \caption{FCT and goodput on the parallel network topology at various loads, when NegotiaToR is equipped with informative requests.}
        \label{table:informative-equests}
        \end{table}

To further improve goodput and FCT performance, we explore two approaches that utilize informative requests. One goodput-oriented approach is to include aggregated data size of the per-destination queue in requests, potentially improving link utilization and goodput by first scheduling the pairs with more demands. 
Another FCT-oriented approach involves including the waiting time of the head-of-line packets from the per-destination queues into requests, prioritizing the pairs with longer waiting delays, potentially avoiding starvation, and reducing tail FCT. 
Naturally, these improvements come at the cost of additional complexities, such as accessing per-destination queue sizes, logging per-packet delays, and implementing sorting mechanisms. 

We implement the two strategies through simulations. For the goodput-oriented data-size approach, we directly include the aggregated data size in requests. While for the FCT-oriented HoL waiting time approach, in the context of priority queues, we need to avoid the long waiting time of elephant flows masking that of mice flows. Therefore, we give the HoL waiting time of the highest-priority queue a greater weight. For instance, based on the three-priority queue case used in our evaluation (\S\ref{sec:evaluation_setup}), here we set the weighted HoL waiting time to be $HoL = (1-\alpha) \frac{HoL_{queue_0} + HoL_{queue_1}}{2} + \alpha \cdot HoL_{queue_2}$, where $queue_2$ is the lowest-priority queue and stores the data of elephant flows.
Our simulations find that the best performance is attained when setting $\alpha$ to a small, non-zero value such as 0.001. This configuration allows for the prompt scheduling of source-destination pairs with mice flows, while also indicating the transmission needs of elephant flows.

\vspace{6pt}

\noindent{\textbf{Simulation results.}} We do simulations on the parallel network topology, and show the results in Table \ref{table:informative-equests}.
The data-size approach attains extremely minor improvements in goodput, while adversely affecting FCT. The HoL-delay approach reduces FCT by 29.9\% at 100\% load, but offers minor to zero improvements at other loads. We thus conclude that the benefits of informative requests are not significant enough to justify the additional complexity. Binary requests, along with the scheduling delay bypassing mechanism, are sufficient to achieve high performance.

\begin{table}[t]
    \resizebox{\columnwidth}{!}{
    \begin{tabular}{|l|ccccc|}
    \hline
    \multirow{2}{*}{}                                             & \multicolumn{5}{c|}{99p Mice Flow FCT ($\mu$s) / Normalized Goodput}                                                                                   \\ \cline{2-6} 
                                                                  & \multicolumn{1}{c|}{10\%} & \multicolumn{1}{c|}{25\%} & \multicolumn{1}{c|}{50\%} & \multicolumn{1}{c|}{75\%} & 100\% \\ \hline
    Base                                                          & \multicolumn{1}{c|}{15.3/9.1\%}     & \multicolumn{1}{c|}{15.4/22.6\%}     & \multicolumn{1}{c|}{15.6/45.2\%}     & \multicolumn{1}{c|}{16.3/67.5\%}     &   22.0/89.0\%    \\ \hline
    Stateful & \multicolumn{1}{c|}{13.5/9.1\%}     & \multicolumn{1}{c|}{13.7/22.6\%}     & \multicolumn{1}{c|}{13.9/45.2\%}     & \multicolumn{1}{c|}{16.3/67.5\%}     &   23.2/88.8\%    \\ \hline
    \end{tabular}
    }
    \vspace{0.1in}
    \caption{FCT and goodput on the parallel network topology at various loads, when NegotiaToR maintains traffic matrices for stateful scheduling.}
    \label{table:stateful-scheduling}
\end{table}

\subsubsection{Stateful scheduling}
\label{sec:stateful-scheduling}

To explore the benefits of stateful scheduling, 
we maintain a stateful traffic matrix at each destination ToR to track incoming traffic demands from all sources, thereby preventing over-scheduling and reducing bandwidth waste. After data arrives at the source, the source ToR sends requests to the destination ToR with the size of newly arrived data to update the matrix. 
Grants are given only if the matrix indicates that the source still has pending data to send, preventing unnecessary scheduling. After giving grants, the size of the remaining data to send in the matrix is decreased temporarily. After the source ToR gives accepts, the accept result will be piggybacked to the destination, allowing for matrix updates. If the source rejects the grant, the destination will revert the temporary adjustment of the matrix to accurately represent pending data for further scheduling. 

\vspace{6pt}

\noindent{\textbf{Simulation results.}}
We conduct simulations of NegotiaToR with stateful scheduling enabled on the parallel network topology, and present the results in Table \ref{table:stateful-scheduling}. The results verify our expectations. With stateful scheduling, NegotiaToR's overall performance roughly remains the same. Therefore, we keep the design that the sources will send requests to the destinations, as long as currently there is pending data to send.

\begin{table}[t]
    \footnotesize
    \resizebox{\columnwidth}{!}{
    \begin{tabular}{|l|ccccc|}
    \hline
    \multirow{2}{*}{}                                             & \multicolumn{5}{c|}{99p Mice Flow FCT ($\mu$s) / Normalized Goodput}                                                                                   \\ \cline{2-6} 
                                                                  & \multicolumn{1}{c|}{10\%} & \multicolumn{1}{c|}{25\%} & \multicolumn{1}{c|}{50\%} & \multicolumn{1}{c|}{75\%} & 100\% \\ \hline
    Base                                                          & \multicolumn{1}{c|}{15.3/9.1\%}     & \multicolumn{1}{c|}{15.4/22.6\%}     & \multicolumn{1}{c|}{15.6/45.2\%}     & \multicolumn{1}{c|}{16.3/67.5\%}     &   22.0/89.0\%    \\ \hline
    ProjecToR & \multicolumn{1}{c|}{16.3/9.1\%}     & \multicolumn{1}{c|}{21.6/22.6\%}     & \multicolumn{1}{c|}{40.8/45.0\%}     & \multicolumn{1}{c|}{52.2/66.1\%}     &   54.4/84.7\%    \\ \hline
    \end{tabular}}
    \vspace{0.1in}
    \caption{FCT and goodput on the parallel network topology at various loads, when we utilize ProjecToR's \cite{projector} scheduling algorithm.}
    \label{table:projector}
\end{table}

\subsubsection{Other design choices}
\label{sec:other-design-choices}
We also explore other design choices, including the distributed scheduling algorithm used in ProjecToR \cite{projector}. ProjecToR also aims to schedule the optical links among multi-ported ToRs in an on-demand manner. Even though it works in a free-space optical switching setting, it is still instructive to transfer its scheduling algorithm to our scenario and compare it with NegotiaToR Matching.

ProjecToR utilizes iterative request-grant based scheduling. It requires measuring and logging data's waiting delays at sources to determine the scheduling priority. Data is grouped into bundles, where each bundle is the unit of scheduling \footnote{In our context, this is the amount of data that can be sent in one epoch.}. Bundles with higher waiting delays are scheduled first. Meanwhile, requests are per-port level, which means, unlike NegotiaToR, when sending requests, the sending port of corresponding data is already chosen.
We transfer it to NegotiaToR, replacing NegotiaToR Matching, and evaluate its performance through simulations.
For comparison, we set the bundle size to be the same with the amount of data sent in one epoch, and only run one round of iteration. Other designs, like NegotiaToR's incast-optimized scheduling delay bypassing scheme, including the priority queue for mice flow prioritization, remain the same.

\vspace{6pt}

\noindent{\textbf{Simulation results.}} We show the results in Table \ref{table:projector}. Even though introducing more complexity, like measuring and logging data's waiting delays, ProjecToR's performance is still inferior to NegotiaToR, both in FCT and goodput. Adding iterations back to ProjecToR can potentially improve goodput, but will also increase the scheduling delay and thus further enlarge mice flow FCT. 
This highlights the effectiveness of NegotiaToR's minimalist design.

\subsection{Micro-observation of NegotiaToR's performance under incast and all-to-all workloads}
\label{sec-appendix3}

\begin{figure}
    \centering
    \includegraphics[width=0.8\linewidth]{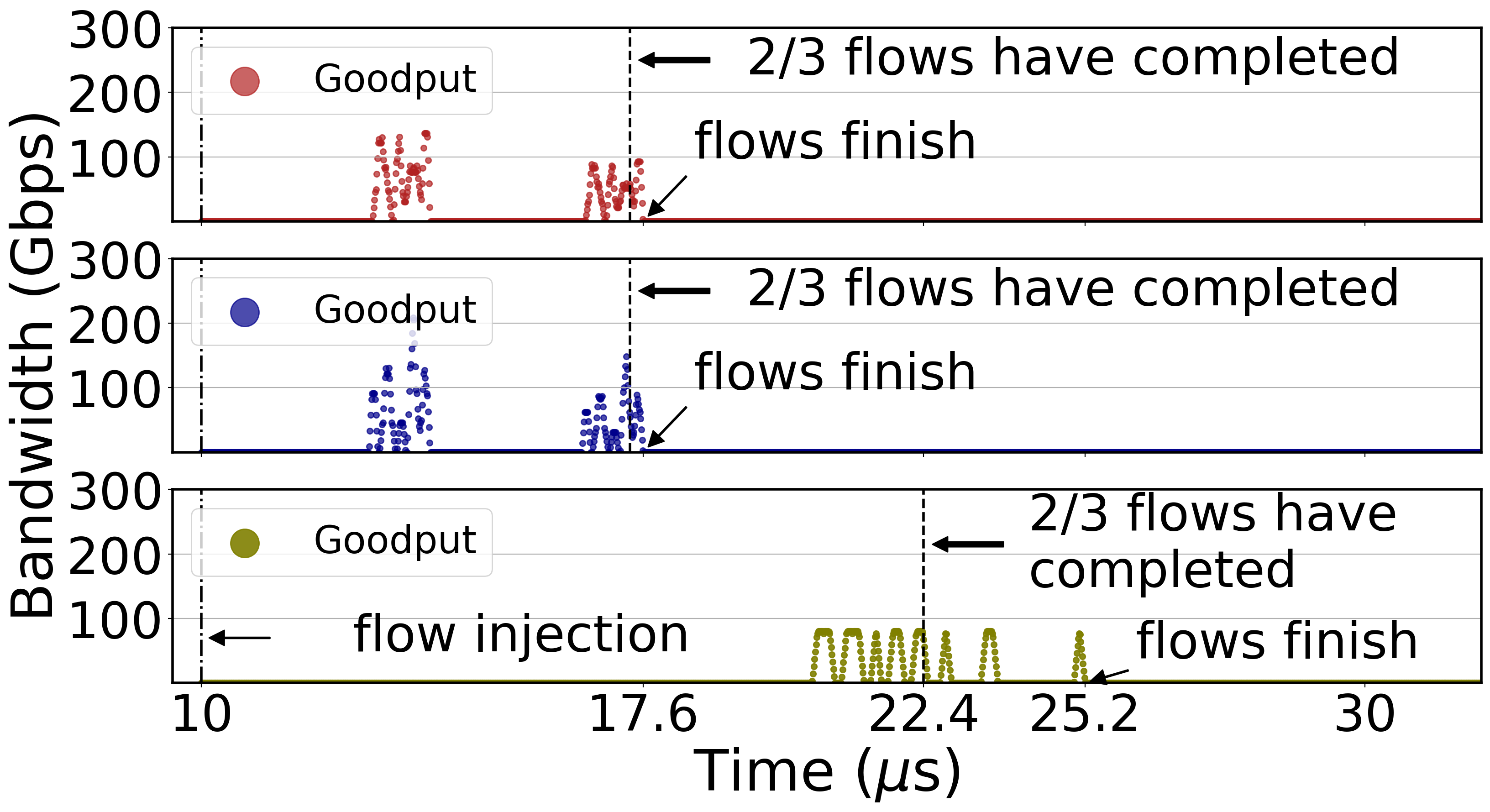}
    \caption{Incast workload: receiver bandwidth observation of incast degree 15. From upper to lower is the bandwidth of NegotiaToR on parallel network, NegotiaToR on thin-clos, and traffic-oblivious scheme on thin-clos.}
    \label{fig:motivation_incast}
\end{figure}

\noindent{\textbf{Micro-observation of the incast workload.}} In the microbenchmark in \S\ref{sec:evaluation_micro}, we tested the performance of incast workloads. To gain further insights, here we sample the bandwidth usage of incast degree 15 at the destination side, as shown in Figure \ref{fig:motivation_incast}.
Flows are injected at 10 $\mu s$. For traffic-oblivious designs, a long interval exists before the destination receives data, because the data needs to be relayed to a third ToR first. 
Whereas for NegotiaToR, due to the incast-optimized scheduling delay bypassing scheme, data can be promptly sent in the predefined phase, so that the destination receives data shortly after the injection, contributing to a short incast finish time. NegotiaToR performs identically across the parallel network and the thin-clos topology, because both topologies provide the same predefined phases.

\begin{figure}
    \centering
    \includegraphics[width=0.8\linewidth]{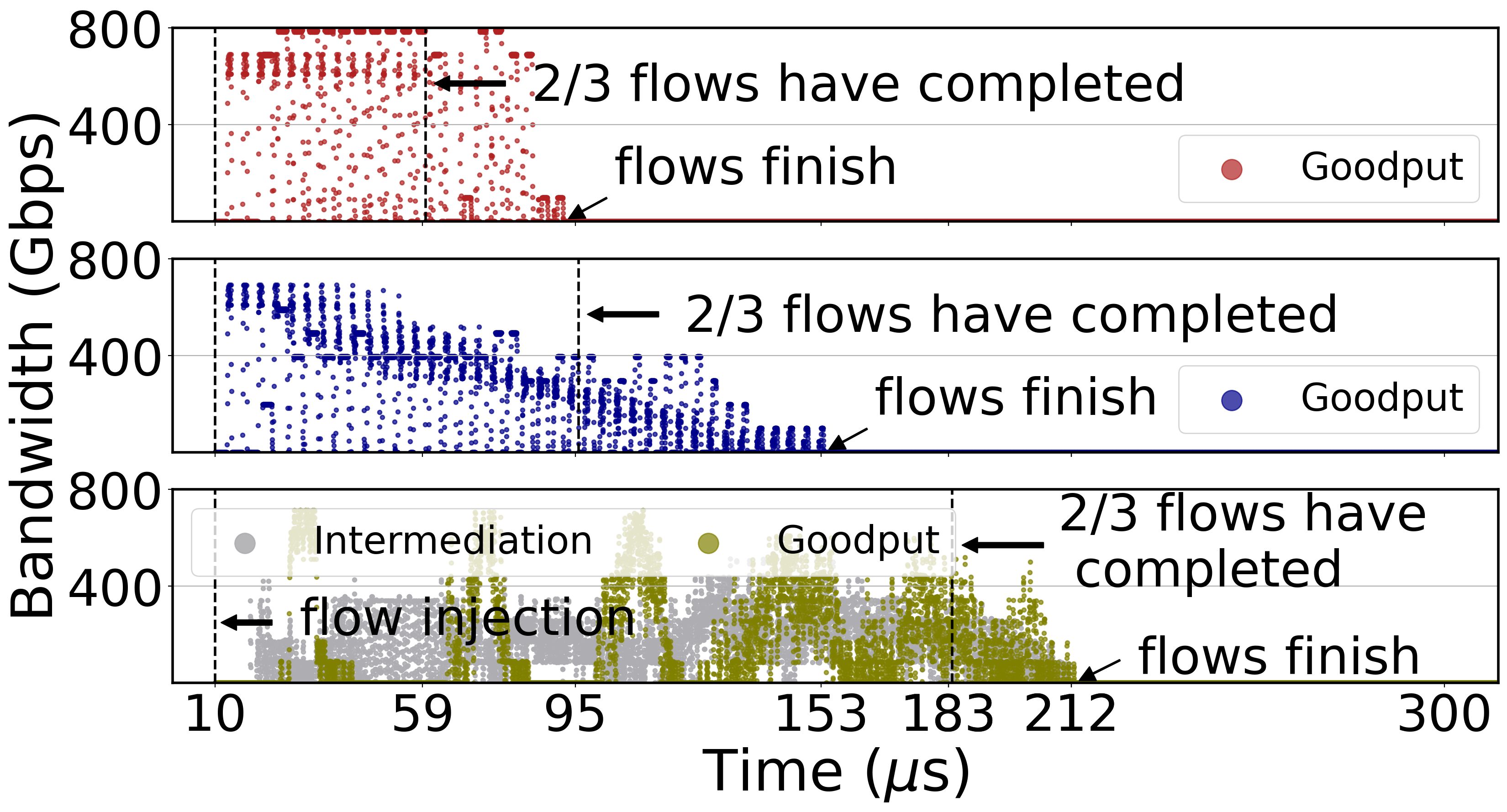}
    \caption{All-to-all workload: receiver bandwidth observation of all-to-all flow size $30KB$. From upper to lower is the bandwidth of NegotiaToR on parallel network, NegotiaToR on thin-clos, and traffic-oblivious scheme on thin-clos.}
    \label{fig:motivation_permutation}
\end{figure}

\vspace{6pt}

\noindent{\textbf{Micro-observation of the all-to-all workload.}} For the all-to-all workload in \S\ref{sec:evaluation_micro}, at a randomly chosen destination, we sample the bandwidth usage when the flow size 30 KB, as shown in Figure \ref{fig:motivation_permutation}. Flows are also injected at 10 $\mu s$. For traffic-oblivious solutions, different from the case of incast workloads, other than the data destined to it, it also receives intermediate traffic (plotted as light-grey dots) that needs to be forwarded, which competes for the receiver bandwidth and does not contribute to goodput of this receiver.
Meanwhile, for NegotiaToR, all received traffic is desired by this destination. As a result, with NegotiaToR Matching's on-demand scheduling, the destination in NegotiaToR continuously receives data at a high bandwidth, leading to high overall goodput.

\subsection{Micro-observation of NegotiaToR's behavior under link failures}
\label{sec-appendix2}

\begin{figure}
    \centering
    \includegraphics[width=0.8\linewidth]{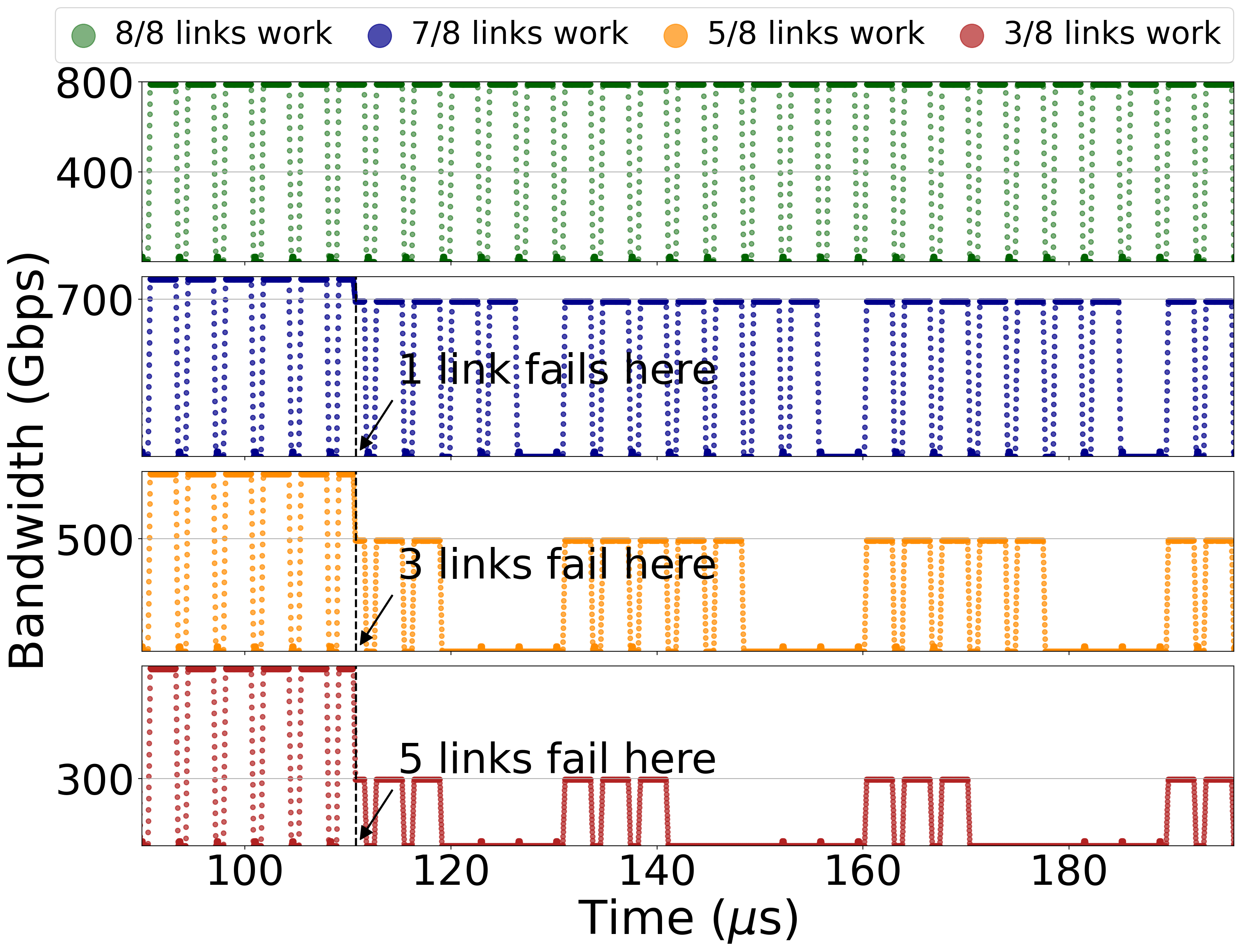}
    \caption{Fault tolerance micro-observation: bandwidth occupation after link failures.}
    \label{fig:fault_micro}
\end{figure}

To better understand NegotiaToR's behavior under link failures, we let a source-destination pair continuously transmit data, and observe the changes of the receiver side bandwidth occupation after link failure happens on the parallel network topology, as shown in Figure \ref{fig:fault_micro}.

When all links work, the bandwidth occupation shows an on-off shape, indicating different epochs. 
With links failing, the bandwidth occupation drops to the level of the remaining links. 
Meanwhile, note that there are epochs with zero bandwidth occupation. 
This is because of the loss of scheduling messages caused by link failures. 

When the source-destination pair happens to rely on one of the failed links to send scheduling messages, the source will not receive any grants, and thus data transmission will be suspended (i.e., all epoch's bandwidth occupation is zero), leading to severe goodput reduction. However, since we regularly change the round-robin rule of the predefined phase, each source-destination pair will send scheduling messages through different links consecutively, mitigating the risk of over-reliance on a single link. 
As a result, this source-destination pair can still transmit data through other links, and the bandwidth occupation is not continuously zero.

\end{document}